# Positivism in Newtonian Mechanics: The Ousia or a Historical Liability?

Seyed Mohammad Rezaei Niya[*]

## Abstract

The positivistic assumptions of determinism and objectivism in the realm of Newtonian mechanics are questioned in this paper. While objectivism is only challenged through proposing the mildest form of subjectivism, determinism is structurally disputed by proposing that the physical reality, at least in the examples discussed, is, in essence, probabilistic and unpredictable. It is discussed that the physical reality and experimenter's identification of it could basically have inconsistent characteristics, and the study of the physical reality can therefore be conducted in ontic and epistemic levels, leading to two distinct identifications. Four scientific topics, showing two different types of indeterminacy, are introduced and briefly reviewed: chaotic systems, turbulence, fluid transport in porous media, and hydromechanics of fractures. It is proposed that determinism is only meaningful in epistemic level, the first two examples are ontically indeterministic, the last two examples are epistemically indeterministic, and more examples of indeterministic phenomena could, most likely, be found in the nature. Indeterminacy of the physical reality, it is discussed, has always been considered in engineering design processes and such effects have normally been covered through safety factors and feedback loops. By reviewing Hadamard's well-posedness criteria, Poincaré's complete deterministic approach, and Leibniz's principles of sufficient reason and identity of indiscernibles, it is claimed that positivism stands on Leibniz's metaphysical assumptions, which are not necessarily in full agreement with the physical reality. A few suggestions for a path beyond positivism in Newtonian mechanics are finally provided.

## Acknowledgements

The discussions with Dr. Serveh Naghshbandi have substantially contributed to the elaboration of the concepts presented. She has also reviewed and edited the text, which is gratefully acknowledged.

## Statements and Declarations

No funding was received for conducting this study. The author has no financial or proprietary interests in any material discussed in this article. This work has not been published previously and is not under consideration for publication elsewhere.

---

[*] Independent scholar, rezaeiniya@gmail.com, seyed.niya@rmit.edu.au

*Barbarus hic ego sum, quia non intelligor illis*, Ovid, Tristia

# 1. Introduction

Despite the significant challenges facing positivism in the last century or so, Newtonian mechanics has been a safe and sacred arena for its assumptions and principles. The sovereignty of positivism, as the only valid paradigm in this realm, is so venerable that only a handful of related published literature can be dug out of the literature on this subject, and those are either on the affirmative side (e.g., Solari and Natiello 2019), or with political considerations (e.g., Alam 1978)[1].

Due to exceptionally wide range of subjects needed to be discussed in the following, previous bitter encounters of misreading in interdisciplinary subjects (e.g., see the appendix in Rezaei Niya et al. 2021), and since "[s]cientists have not generally needed or wanted to be philosophers" as well mentioned by Kuhn (1962, 88), I will employ more-than-(my)-usual-descriptive, sometimes simplistic, but always as accurate as possible[2] language throughout this text. I am also hopeful that radical critiques of reading styles of the scientists[3] are not authentic descriptions for the readership of this text.

First, I need to present my descriptive interpretation of "Newtonian mechanics" and "positivism". I employ "Newtonian mechanics" more as an exclusionary term to emphasise that no quantum-mechanics-related or relativity-related interpretation or consideration is applied. Fluid mechanics, solid mechanics, and chaotic systems are branches of specific interest. Here, I also exclude any thermodynamical, microscopic, entropy-related phenomenon or representation[4]. By "positivism", in a simplistic interpretation, however, I include all the hidden assumptions, primarily philosophical and some metaphysical which will be discussed in the following, that scientists in various Newtonian-mechanics-based branches of science, employ in their regular scientific publications[5]: e.g., that layers of identity, prior knowledge and experiences, and in general, subjectivity of the experimenter are not involved in the findings reported (objectivist assumption); that the article published, if well written, is self-sufficient and self-explanatory (phenomenological assumption); and that the experiments conducted, if well documented and reported, can be indefinitely repeated in similar environments everywhere and any time in the world (deterministic assumption)[6]! Many more hidden assumptions can be listed here. The invitation to reflect on accuracy of the presented assumptions, while points of deep concerns in Social Sciences (e.g., O'leary 2004; Phillips and Burbules 2000; Robson and McCartan 2016), might be considered far-fetched inside Natural Sciences (although, not every scientist-philosopher supports this idea, see Polanyi 1962). My critique to positivism in the following will be, at least pragmatically, much more benign; however, the deep-seated faith in and trust to positivism is so common and widespread between natural scientists and philosophers of science[7] that even mild questioning of positivistic assumptions could lead (and have led) to "science wars" (e.g., see Bloor [2004] on one side, and Zammito [2004] on the other).

---

[1] While the critiques provided by Polanyi, Kuhn, and many others around the mid-20th century and before, and the scholars of Sociology of Scientific Knowledge (SSK) in recent decades raise fundamental questions on inherently-positivistic assumptions about the character of scientific progress, they do not challenge positivism from *within* Newtonian mechanics; e.g., objectivism is not necessarily challenged through such discussions.

[2] I am aware that such positivistic phrases, or passive sentences in present tense, employed here and there in the text, the consequence of my natural scientific training and work experience, might not please the philosophers and epistemologists. I recommend these respected readers to consider the hidden subjective suffixes "for me", "by me", "to me", etc. in such phrases and sentences.

[3] "You have to realize that most physicists don't read. Reading is not part of our culture. When we see a book or a paper we dip into it here and there and make a decision… It can [also] lead to serious trouble in attempts to assess less byzantine texts whose authors are, however, labouring under the illusion that they are addressing a reader who is following everything that is being said.", Prof. David Mermin, Cornell physicist, quoted in (Collins and Pinch 1998, 153).

[4] Such phenomena could bring other types of obstacles in front of determinism and positivism, as discussed by Polanyi (1962, 412-415).

[5] For a more detailed discussion and definition of positivism, the readership is referred to research methodology handbooks; e.g., O'leary (2004), Phillips and Burbules (2000), and Robson and McCartan (2016).

Interestingly, no critique of positivistic assumptions has been presented *within* Newtonian mechanics, to the best of my knowledge[8]. While inaccuracy or impossibility/impracticality of some assumptions, directly resulted from positivism, belongs to the *common knowledge* of engineers in some branches of Newtonian mechanics (specifically, fluid mechanics), rephrasing this common knowledge to reveal its roots in some *metaphysical* assumptions, disguised in positivism, normally leads to unexpected reactions (see the appendix in Rezaei Niya et al. 2021). In this regard, this paper is only an attempt to demonstrate that specific types of research questions, some have long been abandoned but some are still persistently followed, are in essence ill-defined due to the positivistic assumptions which are not in agreement with physical realities.

---

[6] I hope these extreme examples are not used as counter-attacks against the critiques directed to positivism in this text!

[7] Natural scientists are, for the most part, unfamiliar with the related philosophical discussions and such analyses are notably absent in recent literature (see the appendix in Rezaei Niya et al. 2021). On the other hand, philosophers of science come primarily from analytic philosophy tradition, in which positivism is indeed considered *the Ousia*. Regarding Comte's positivism, John Stuart Mill (1973, 137) states: "We, therefore, not only hold that M. Comte was justified in the attempt to develop his philosophy into a religion,… but that all other religions are made better in proportion as, in their practical result, they are brought to coincide with that which he aimed at constructing."
Professional philosophers are, in general, not well-known for their enthusiasm towards embracing new ideas (Jacoby 1987, Ch. 6). Philosophers of science either overlook nearly a century of discourse on the onto-epistemological position of positivism (such questionings had already been started by Veblen [1918, Ch. 1]) and reduce positivism mainly to logical positivism (While even Encyclopedia Britannica has an entry on "positivism", Stanford Encyclopedia of Philosophy does not have such an entry and only covers an entry on "logical empiricism", which states [Creath 2023] that logical empiricism "has no very precise boundaries and still less that distinguishes it from 'logical positivism'."), or reject altogether other ontological and epistemological assumptions (International Encyclopedia of the Social & Behavioral Sciences's entry on positivism [Turner 2001] starts with "Positivism is the name for the scientific study of the social world". Regarding the references cited in footnote 5, one reviewer once stated those are "a bogeyman found in some non-philosophical treatments of methodology").

[8] The literature of philosophy of science is not unfamiliar with philosophical "thought" experiments questioning determinism even in classical mechanics. Earman (1986) has reviewed a list of such examples, only to conclude that "[c]lassical physics would seem to be a poor choice of hunting grounds for such examples [of divine intervention or accident] since, as we all know, the laws of classical physics are deterministic in the Laplacian sense. We know no such thing, at least if knowledge implies truth" (p. 23), or more specifically, "[t]he special theory of relativity rescues determinism from the main threat it faces in Newtonian worlds, and in special relativistic worlds pure and clean examples of determinism, free of artificial props, can be constructed." (p. 2). On the other hand, another category of "thought" experiments, known as Lipschitz-indeterminism, has been re-surfaced in the literature of philosophy of science in recent years (e.g., Norton 2003, 2008; Malament 2008; Korolev 2007; Zinkernagel 2010), while it has been known in the mathematical literature for more than a century (van Strien 2014). The Lipschitz continuity criterion (named after Rudolf Lipschitz, a well-known nineteenth century German mathematician) as a part of Cauchy-Lipschitz theorem for the existence and uniqueness of the solution of an initial-value problem can be found in any classical reference book on differential equations (e.g., Petrovskii 1966) and goes back to the nineteenth century (Lindelöf 1894).
The principal figures of positivism and determinism cited in this text, Poincaré, Hadamard, and Wittgenstein, were all prominent mathematicians of their time, and, as one could expect, were well aware of such anomalies. The Hadamard's well-posedness criteria and discussions quoted from him regarding Cauchy's problem in footnote 25, and Wittgenstein's emphasis on falsehood of a proposition depending on a non-existent complex, discussed in section 6, are specifically attacking such forms of "thought" experiments.

I question causal determinism in this article as the cornerstone of positivism. I propose that the physical reality, at least in the four topics discussed in the following, is in essence probabilistic and unpredictable, in the scales much larger than any quantum-based uncertainty principle and for fundamentally different reasons. Reviewing implied methods of implementing determinism in scientific investigations, I discuss Hadamard's well-posedness criteria, and argue that Leibniz's principles of sufficient reason and identity of indiscernibles as the foundations of causal determinism are primarily based on Leibniz's metaphysical assumptions. In addition, I claim that at least the weakest form of subjectivity is needed to be considered in the realm of Newtonian mechanics; i.e., that the physical reality and the scientist's identification of it could basically have different characteristics.

The four scientific topics, fitting perfectly into my narration, are first briefly reviewed: chaotic systems, turbulence, fluid transport in porous media, and fractures. The ontic and epistemic identifications of the physical reality, the entry point of subjectivism into the realm of Newtonian mechanics, are introduced, and the Hadamard's well-posedness criteria are discussed from this perspective. The further physical realities that potentially follow indeterminism are presented, and the classic approaches to deal with indeterminacy[9] in design processes, again common knowledge of any engineering designer, are reviewed. It is then discussed that determinism, hidden in the Hadamard's well-posedness through his reference to Poincaré, is founded, at least in the enlightened world, on Leibniz's metaphysical assumptions/analyses. The well-known notion of "deterministic, but unpredictable" (chaotic) systems is debated. It is finally proposed that meaningful room for post-positivistic assumptions in Newtonian mechanics needs to be considered.

## 2. Counterexamples of determinism

### 2.1 Chaotic systems

Chaotic behaviour has been experienced and studied in many different fields as diverse as biology, physics, chemistry, and various disciplines in engineering. Chaos theory has become a celebrated topic in recent decades and been grown to a distinct branch of science. A rich body of scientific, philosophical, and even popular science writings in different levels has been developed around chaos (see, for example, Bishop 2017; Cencini et al. 2021, Ch. 7; Gleick 1987; Strogatz 1994). Nevertheless, a precise definition of chaos or even chaos theory is still lacking (Bishop 2017). A simplified definition of a chaotic system, at least useful for the following discussions, is a dynamical system with an unpredictable behaviour due to (at least) extreme sensitivity to initial conditions, which itself stems from the nonlinearity of its governing equations. While presenting complex behaviours potentially leading to various types of attractors, a chaotic system can be developed from an equation as simple as a nonlinear recurrence relation, discussed in Section 4.2.

---

[9] The word "indeterminacy" has been occasionally employed in philosophy of science literature to indicate the consequences of Heisenberg's uncertainly principle (e.g., Hoare 1932). Since the scope of this paper is limited to Newtonian mechanics, "indeterminacy" should be interpreted here only as a noun indicating the types of "indeterminism".

The extreme sensitivity to the initial conditions leading to unpredictability was known at least from Poincaré's studies on the three-body problem (Cencini et al. 2021, Ch. 22). Lorenz's study in the 1960s on simplified convection equations, however, is normally cited as the cornerstone of the chaos theory (see, e.g., Bishop 2017; Cencini et al. 2021, Ch. 7)[10]. For the purpose of this paper, it is sufficient to emphasise the unpredictability of a chaotic system in any meaningful way, leading to a blurry image of the future of the related physical reality, which is widely accepted by scholars of the field, even vigorous advocates of positivism, like Poincaré himself. It is stated that expecting a qualitative prediction of the system (Kellert 1993) with a pinch of quantitative approximation (Smith 1998) and statistical information (see Bishop 2017) is the most that one can expect from the analysis of a chaotic system, at least to our current understanding.

## 2.2 Turbulent flow

Most of fluid flows experienced in the nature have characteristics normally categorised them as "turbulent flows" in fluid mechanics. The word "turbulence", coming from "turbolenza" in Italian, was presumably employed first by Leonardo da Vinci (Colagrossi et al. 2021) to describe the complex nature of this specific type of flow, which is considered "the last great unsolved problem of classical physics" (Holmes et al. 2012, 3). Contrary to the laminar (layered) flow, in which the flow is by any definition deterministic[11], the unpredictability of instantaneous properties (e.g., velocity, pressure, temperature) of turbulent flow in any point inside its domain can be considered the common denominator of various definitions presented for this phenomenon. Reynolds (1883), who first quantified the transition from a laminar to a turbulent flow, identified turbulence as sinuous and "irregular"[12], and stated that Stokes has pointed out in 1843 that under certain conditions, the motion becomes unstable and sensitive to "indefinitely small disturbance". Taylor (1920) resembled turbulent flow to "molecular agitation" and "random migration". The "irregularity" and "randomness" gradually became the permanent adjectives for turbulence in the literature (Taylor 1935; von Kármán 1937; Dryden 1939; Hinze 1959; Hunt et al. 2001; White 2009; Schlichting and Gersten 2017) and statistical methods were developed and are still being employed to analyse the average properties of turbulent flows. Turbulence has also been studied as a chaotic system and its coherent structures have been mainly investigated in recent decades (Holmes et al. 2012).

---

[10] It is a potential research inquiry for sociologists of scientific knowledge that how much popularity of the chaos theory was due to the questioning of positivistic conception of scientific progress raised by Kuhn (1962), Polanyi (1962), and many others.

[11] assuming the initial and boundary conditions are deterministically specified.

[12] Reynolds (1883, 1894) called laminar and turbulent flows "direct" and "sinuous" motions, respectively. Taylor also discussed "sinuosity" of a wind curve in his 1914 paper. It was first in his 1920 paper that Taylor used the "turbulent motion" phrase to cite his 1914 work and the works of others.

Either stochastic-like fluctuations superimposed on a predictable mean flow (Schlichting and Gersten 2017), chaotic-like oscillations leading to coherent structures (Holmes et al. 2012), or even some complex combination of both (Holmes et al. 2012, 65-66), turbulent flow, according to a solid consensus in the related literature, is instantaneously unpredictable. On the other hand, it is strongly believed that turbulent flow is adequately described by Navier-Stokes equations (Holmes et al. 2012, 19). While, interestingly, not normally considered a challenge to determinism, turbulence can be claimed to be a "deterministic, but unpredictable" phenomenon, similar to any other chaotic system[13].

**2.3 Fluid transport in porous media**

The unpredictability of the first two examples, introduced above, are well accepted in the literature, and no paper is, for example, presented now and again claiming a new correlation to predict the instantaneous temporal or spatial velocity pattern of a turbulent field. These scientific topics are commonly considered "unpredictable, but deterministic".

The next two examples, however, are not even presumed "unpredictable" on the surface level of the related literature, and it is only deep in the details of the analyses in the specialised papers and books that the fundamental obstacles in quantification of such phenomena emerge; I will call this category indeterminacy Type II, while the first two examples, discussed in Sections 2.1 and 2.2, are categorised as Type I. Active research on these topics is still in progress, and further correlations, theories, and "universal laws" are occasionally claimed, always with limited success, after close to two centuries of extensive and multidisciplinary research, often with a rich mathematical literature, at least for the two examples discussed in the following.

Fluid transport[14] in porous media has been thoroughly studied in various engineering disciplines, at least from Darcy's 1856 pioneering work (Brown 2002), due to its extensive applications in numerous engineering fields (Rezaei Niya et al. 2021). A quick glance at the related references (e.g., Bear 1972; Bear and Cheng 2010; Ichikawa and Selvadurai 2012; Nield and Bejan 2006; Scheidegger 1974) reveals that hundreds of papers have been published on this topic, and many elaborate analytical, computational, and experimental techniques and approaches have been developed. The very basic question of how permeability, the fluid transport capability, of a porous medium can be accurately determined is, however, still open. Contrary to above-mentioned examples (i.e., indeterminacy Type I), it does not mean that permeability cannot be accurately *measured* for a specific porous

---

[13] Direct Numerical Simulation (DNS) of turbulence (e.g., Alfonsi 2011), accurately predicting the statistical characteristics of turbulent flows in limited cases (due to its extreme computational expense), could lead to the illusion of turbulence determinability. Since I deal with similar cases in detail in the following sections, I only need to emphasise here that from the determinism point of view, DNS of turbulence is in essence equivalent to modeling of (laminar) fluid transport in porous media and fractures (indeterminacy Type II), and any critiques presented to the latter are applicable to the former too.

[14] specifically, pressure-head-driven advective transport, even at low-Reynolds-number creeping flow regime (see Rezaei Niya et al. 2021)

structure, or even accurately *predicted through computer modelling* (at least, theoretically) when the porous structure is fully obtained[15]. However, the permeability *prediction without outright measurement* is, at least so far, only possible with a limited accuracy[16]. To do so, the porous structure needs to be somehow quantified by (a limited number of) parameters that can accurately represent such a structure, and the attempts towards such an accurate identification have been so far unsuccessful[17].

Trained in a positivistic culture and paradigm, each generation of researchers, including the author[18], has conducted new attacks to this problem and contributed to the already vast body of available literature (e.g., see Menke et al. 2021; Xiao et al. 2022, and the references cited in Rezaei Niya et al. 2021). The literature accepts that accurate estimation of permeability is not possible without a detailed identification of the porous structure, and also "a detailed description of the pore space is impossible" for most applications (Bear 1972, 41); but the direct conclusion of impossibility of accurate estimation of the permeability of a general porous structure, and its resultant questioning of the positivistic assumptions, are, interestingly, passionately rejected (see the discussions provided in the appendix in Rezaei Niya et al. [2021])[19].

## 2.4 Fractures

While might not be clear in the first sight, fluid transport in fractures is fundamentally similar to transport in porous media, and the leading scholars of the latter are normally the influential authors of the former too (e.g., Bear et al. 1993; Scheidegger 1963; Selvadurai and Suvorov 2017). The contact of two rough surfaces with or without a liquid in between, as a physical reality, and its mechanical characteristics appear, however, in much wider engineering applications than only underground construction, excavation or fluid transport.

The hydro-mechanical characteristics of fractures are currently under active research, specifically due to their applications in various energy industries (see Rezaei Niya and Selvadurai 2019; 2021). Any accurate estimation of either fluid transport or mechanical response depends on an accurate identification (and reproduction for modelling purposes) of fracture surfaces. Similar to transport in porous media, if the fracture surfaces can be accurately determined, the fluid transport and mechanical performance can be accurately predicted, at least in theoretical level[20]. The reproduction or quantification of fracture surfaces, however, have been proven to be a daunting task, and employing even up to 20 parameters to quantify a fracture surface was not necessarily successful (Smith 2014). On the other hand, the statistical analysis of fracture surfaces, quite prevalent in the literature, leads to estimating a range for fracture response rather than an accurate prediction (see Rezaei Niya and Selvadurai 2019; 2021). Nonetheless, it does not mean that universal laws are not claimed occasionally in the literature (e.g., Petrovitch et al. 2013; Pyrak-Nolte and Morris 2000; Pyrak-Nolte and Nolte 2016; Wang and Cardenas 2016) with limited applicability[21].

---

[15] Some serious concerns on the accuracy and applicability of such visualisation techniques for vast categories of natural and artificial porous media have been raised, see Rezaei Niya et al. (2021).

[16] with sometimes up to two orders of magnitude error, see Rezaei Niya and Selvadurai (2017; 2018) and Rezaei Niya et al. (2021)

[17] I have discussed in detail the literature of such attempts, the parameters defined (specifically, porosity and tortuosity of the porous structure), and the inevitable inaccuracies raising from employing such parameters in Rezaei Niya et al. (2021).

[18] My research experience in the last two decades has mostly been on mathematical and computational modeling of various engineering systems (see Tabrizi et al. 2004; Aghdam et al. 2006; Norouzi et al. 2012; Rezaei Niya et al. 2010; 2014; 2015; 2016; 2016-2; 2022). During my research on fluid transport in porous media in the last couple of years, my position has shifted from attempting to develop new approaches (Rezaei Niya and Selvadurai 2017; Selvadurai et al. 2017) and parameters (Rezaei Niya and Selvadurai 2018) for accurate (deterministic) estimation of the permeability towards accepting the impossibility of such objective (Rezaei Niya et al. 2021), and estimating statistical characteristics of such physical realities instead (Rezaei Niya and Selvadurai 2019; 2021; Selvadurai and Rezaei Niya 2020).

[19] It needs to be emphasised here that the conclusion of indeterminacy could be easily questioned without a clear distinction between ontic and epistemic identifications, discussed in Section 3.

## 3. Ontic versus epistemic identification

Before any further discussion on indeterminacy of the examples presented above, I need to clarify the distinction between ontic and epistemic identifications of a physical reality. While similar, but not exactly identical, conceptualisations and discussions have already been elaborated under scientific representation (Frigg and Nguyen 2021) and/or modelling in science (Frigg and Hartmann 2020), this clarification is specifically important for the following arguments against determinism.

Any physical reality or phenomenon can be investigated in ontic or epistemic level. In ontic level, the phenomenon *per se* is studied. The onset of turbulence in a pipe, prior to introduction of Reynolds number (Reynolds 1883), could be employed as an example here. One could accurately determine the transition to turbulence for various fluids and pipes. The phenomenon is then ontically determined. It, however, could not provide any information about this transition for unmeasured pipes, fluids, or velocities. While various realisations of this phenomenon have been accurately measured, no generalisation or prediction can be achieved. The phenomenon, in essence, has not yet been *understood*.

In the epistemic level, the phenomenon, on the other hand, is studied *in comparison* or *in the context* of similar phenomena, is *defined* presenting *inclusion/exclusion* criteria, is *quantified* and *indexed* employing the limited number of parameters, is *theorised* and *generalised*, and is described drawing on the previously-known or newly-developed theories. The outcome of epistemic identification of a physical reality could be an experimental relation (onset of turbulence in a pipe), a model, a theoretical analysis, or even a qualitative description.

---

[20] The accurate prediction of hydro-mechanical response of a fracture, even when the surfaces of the fracture are perfectly known, has significantly more substantial obstacles than those of fluid transport in porous media, due to local heterogeneity of rock surface materials, limited understanding of asperity breaking process, simplified theoretical models available, and extreme sensitivity of the fracture characteristics to relative surface displacements (see Rezaei Niya and Selvadurai 2019; 2021).

[21] The proposed universal correlations, in the references cited, basically consist of a vertical-line section in which parameter of *y*-axis is independent of the parameter of *x*-axis, and a horizontal-line section in which parameter of *x*-axis is independent of *y*-axis parameter (see Fig. 7 in Wang and Cardenas [2016], and Fig. 4 in Pyrak-Nolte and Nolte [2016])! Similar correlations can be developed for fluid transport in porous media as well (see Fig. 4 and Fig. SM.4 in Rezaei Niya and Selvadurai [2017]).

Particularly, prediction, either accurate or inaccurate, is only possible after epistemic identification[22].

Ontic and epistemic identifications are, clearly, not fundamentally new concepts. Any scientific discussion comprising of modelling, idealisation, and abstract representation is based on epistemic identification. Platonic theory of forms, Aristotelian idealisation (Frigg and Hartmann 2020), and even Galilean idealisation (McMullin 1985) are all emphasising the need for abstraction of a physical reality to achieve an understanding. While ontic identification is an object-oriented endeavour, the epistemic identification should be considered a subject-oriented ideation, normally constructed upon already existing understandings. The familiarity with the specific literature of the problem in hand, the seniority and experience level of the researcher, apprenticeship, and intellectual passion, among many other factors (see Polanyi 1962), could influence mastering epistemic identification.

Having a positivistic perspective, one could suggest that discriminating ontic and epistemic identifications is only "a nice derangement of epistemes"[23]. I argue, however, that they are clearly distinctive, in fact important-to-distinguish, identifications with different characteristics. Table 1 depicts representations of ontic and epistemic identifications of the topics discussed above. The ontic identification column shows *the reality* of a phenomenon for each topic, while epistemic identification column displays *the image* or *the idea* of the same phenomenon. All similar realities have the same image or idea in an all-to-one relation, while this *similarity* itself is normally defined based on the characteristics of the image/idea[24]. The image/idea dictates what parameters of the reality should be considered, and what parameters should be neglected. As an example, the colour of the double-pendulum, or the latitude of the laboratory in which the turbulent flow has been measured, is disregarded. It is normally only a handful of parameters that are transferred from the reality to the image/idea, invigorating the latter to predict the future or past states of the former.

---

[22] One further example could help to differentiate between ontic and epistemic identifications: Suppose a never-seen-before tree with an unusual shape has been discovered; one investigator could study the size, the shape, the number and sizes of the branches, and the shapes and sizes of every single leaf of this tree. The investigator has then performed a thorough *ontic* identification of this tree. They, however, cannot answer some basic questions about this tree: for example, is its unusual shape due to external parameters (e.g., wind direction, physical limitations during growth) or inherent to this type of tree? What is the *typical* shape and size of such a tree type? Is this tree in a *normal* environment for its growth? To answer such questions, an *idea* of this tree type and its environment is required. Such *understanding* and *idea* are only achievable through *epistemic* identification.

[23] borrowing the title of the book by Zammito (2004)

[24] Certainly, many different ideas/images can be developed for a specific reality. This side of the debate is not considered here, see Bishop (2017).

Table 1. Sample representations of ontic and epistemic identifications of the examples discussed

| Examples | Ontic identification | Epistemic identification |
| --- | --- | --- |
| 1. Chaotic system (double-pendulum) | 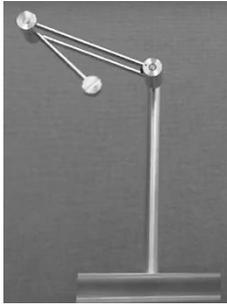 | 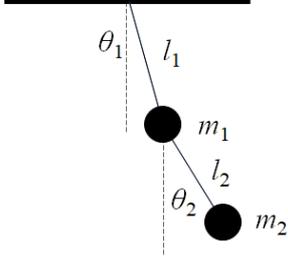 $\theta_1'' = f(\theta_1, \theta_2, m_1, m_2, l_1, l_2, \theta_1', \theta_2')$ $\theta_2'' = g(\theta_1, \theta_2, m_1, m_2, l_1, l_2, \theta_1', \theta_2')$ |
| 2. Turbulent flow | 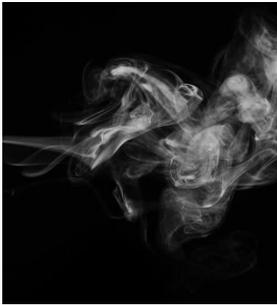 | 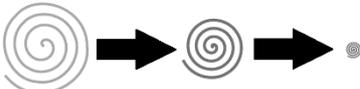 $\rho \dfrac{D\mathbf{u}}{Dt} = -\nabla p + \nabla \cdot \tau + \varrho \mathbf{g}$ |
| 3. Fluid transport in a porous medium | 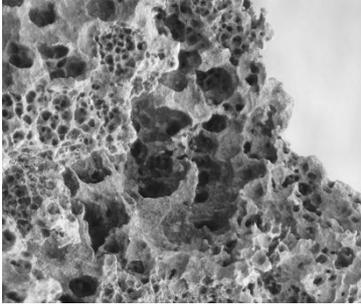 | 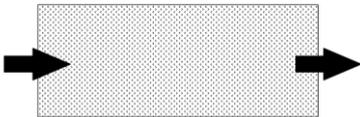 $Q = \dfrac{kA}{\mu L}\Delta p$ |
| 4. Fracture | 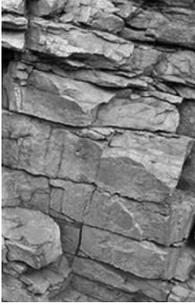 | 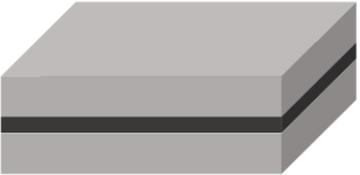 $e_h = \dfrac{e_m^2}{\mathrm{JRC}^{2.5}}$ |

Accuracy level aside, the credibility of a model in scientific applications is traditionally evaluated using Hadamard's well-posedness criteria, according to which a model is *well-posed*, if a) it has a solution, b) the solution is unique, and c) the solution is stable (Hadamard 1902; 1915; 1923). Working on Cauchy's problem, Hadamard, in fact, developed a strong interpretation of positivism, based on his reading of Poincaré (see Hadamard 1902, 49; 1915, 11; 1923, 23, 32, 33, 38), in which not only physical reality (i.e., ontic identification) and the researcher's understanding of it (i.e., epistemic identification) are inherently equivalent, but mathematical abstraction is also bounded by the physical reality[25]. While emerging mathematical attention to *ill-posed* problems (Tikhonov and Arsenin 1977; Tikhonov et al. 1995) rejects, at least, Hadamard's strong positivism[26], the equivalency of ontic and epistemic identifications, as one can guess, can also be questioned.

The Hadamard's own well-posedness criteria are helpful to illustrate the inherent difference between ontic and epistemic identifications. Clearly, for any mathematical equation, including the mathematical representation (i.e., epistemic identification) of any physical reality, the existence, uniqueness, and stability of the solution are important questions to be investigated. However,

a) the existence of the response of a physical reality itself (i.e., ontic identification) is obvious and questioning its existence is meaningless. Each phenomenon, either deterministic or stochastic, has an outcome as a direct consequence of occurring. When a coin is tossed, it does not disappear[27]!
b) the uniqueness of the solution in the ontic level is again obvious, if (absolute) time is considered as an additional dimension. Any stochastic phenomenon even, although may not be predictable, has a unique output in any specific (absolute) time. When a coin is tossed, a head or tail, and not both, is obtained.
c) On the other hand, for Type-II examples discussed above, i.e., fluid transport in a porous medium and fracture response, the physical reality (ontic identification) is sensitive to minute details of the porous structure and fracture surface, respectively, while their epistemic identifications are at least not as much sensitive (see Rezaei Niya and Selvadurai 2017; 2018; 2019; 2021).

Finally, I need to emphasise that any computer modeling of a physical reality, seeking to duplicate the phenomenon in all its details in a simulated reality, should be considered as an ontic identification. While a successful simulation could be judged as the verification of the fundamental equations employed (here, Newton's equations of motion), the simulation, by itself, does not provide any original advancement in *understanding* the physical reality.

---

[25] Hadamard essentially believed that the mathematical problems that do not reflect a physical reality, are not "correctly set". He starts his 1902 paper, quoting the following from Poincaré: "La physique ne nous donne pas seulement l'occasion de résoudre des problems… elle nous fait pressentir la solution". He quotes similar sentences from Poincaré in Hadamard (1915, 11): "It is physics which gives us many important problems, which we would not have thought of without it, … It is by the aid of physics that we can foresee the solutions". He elaborates on this concept in Hadamard (1923, 32), "But it is remarkable, on the other hand, that a sure guide is found in physical interpretation: an analytical problem always being correctly set, in our use of the phrase, when it is the translation of some mechanical or physical question", and p. 38, "If a physical phenomenon were to be dependent on such an analytical problem as Cauchy's for $\nabla^2 u = 0$, it would appear to us as being governed by pure chance (which, since Poincaré, has been known to consist precisely in such a discontinuity in determinism) and not obeying any law whatever". It is interesting that Hadamard directly connects the fate of well-posedness to determinism!

[26] Since these *ill-posed* problems are, in many cases, credible models for specific physical realities, Hadamard's Poincaré-based determinism, according to his own quote mentioned above, consists of discontinuity!

[27] Coin tossing is proposed as an example for a stochastic phenomenon in this text.

Direct Numerical Simulation (DNS) of turbulence, and detailed visualisation of porous structures and fracture surfaces and then computer modelling of fluid flow in those structures[28], even if successful, should, therefore, be considered as ontic identifications. As an example, mere duplication of fluid transport in a specific porous structure in a simulated reality is not inherently different than accurate measuring of the fluid transport characteristics in the physical porous medium from which the simulation has been duplicated. It has also been discussed in the literature before that such analyses *per se* do not lead to understanding of the physical reality (see, for example, Holmes et al. 2012, 4, 7, 37).

## 4. The rationale against determinism

Having introduced Type I (the first two) and Type II (the last two) examples, and distinguished epistemic and ontic identifications, I can now present my rationale against determinism.

### 4.1 Determinism is only meaningful in epistemic level.

A phenomenon is deterministic, only if it can be somehow "predetermined"! Bishop (2017) defines determinability of a model as exhibiting *unique evolution* from a given state. Even the notion of "deterministic, but unpredictable", claimed for chaotic systems, shows that determinability is quite closely associated with predictability. The mere measuring and quantifying a physical reality, either invariant or (absolute-) time-dependent, does not make it deterministic. Knowing the outcome of a tossed coin after tossing, does not make tossing a deterministic process. Similarly, since the permeability of a porous structure can be accurately measured, either directly or through computer modelling of the exact structure in hand, the fluid transport in a porous medium is not necessarily deterministic. This physical reality is deterministic only when permeability can be accurately predicted employing some physical parameters of the medium (e.g., porosity). In other words, it is deterministic only when it is determined in the epistemic level.

### 4.2 Type-I examples are ontically indeterministic.

As will be discussed in Section 6, chaotic systems, from the very beginning, have been considered "deterministic, but unpredictable"[29]. It is traditionally stated that this unpredictability is the result of exponential growth of minutest differences in initial conditions, soon leading to a macroscopically different state of the system. The unpredictability of the system, it is assumed therefore, emerges from impotency of the experimenter, and not from the unpredictability/indeterminability of the physical reality. In other words, the unpredictability of a chaotic system is considered subject-oriented rather than object-oriented.

---

[28] The exact computer reproduction of an already-visualised physical porous structure, for example, is certainly different than computer realisation of porous structures, developed based on some quantifying parameters (e.g., porosity and tortuosity), either flavoured with some randomness or not! While the former is categorised as an ontic identification, the latter is definitely an epistemic identification.

[29] Turbulence has been strikingly missing in the literature of determinism. As an example, Chaos' entry in Stanford Encyclopedia of Philosophy is more than 25,000-words long with a detailed discussion on determinism, but no entry has been provided for turbulence in this encyclopedia!

While any minute difference/error in initial conditions of chaotic systems indeed results in arbitrary large differences in outcome after enough iterations, a chaotic system is also unstable against arbitrary-small truncation errors. To elaborate on this salient characteristic of chaotic systems, normally overlooked at least in philosophical discussions, a simple chaotic recurrence relation, presented in Equation (1), is considered:

$$x_{n+1} = rx_n(1 - x_n) \tag{1}$$

The relation can also be written in an analytic form:

$$
\begin{aligned}
x_1 &= rx_0(1 - x_0) \\
x_2 &= r^2 x_0(1 - x_0)\,(1 - rx_0(1 - x_0)) \\
x_3 &= r^3 x_0(1 - x_0)(1 - rx_0(1 - x_0))[1 - r^2 x_0(1 - x_0)(1 - rx_0(1 - x_0))] \\
x_4 &= r^4 x_0(1 - x_0)(1 - rx_0(1 - x_0))[1 - r^2 x_0(1 - x_0)(1 - rx_0(1 - x_0))] \\
    &\quad \{1 - r^3 x_0(1 - x_0)(1 - rx_0(1 - x_0))[1 - r^2 x_0(1 - x_0)(1 - rx_0(1 - x_0))]\} \\
&\vdots \\
x_n &= r^n x_0(1 - x_0)(1 - rx_0(1 - x_0)) \cdots \\
    &\quad \{1 - r^{n-1} x_0(1 - x_0)(1 - rx_0(1 - x_0)) \\
    &\quad \cdots [1 - r^{n-2} x_0(1 - x_0)(1 - rx_0(1 - x_0)) \cdots ]\}
\end{aligned}
\tag{2}
$$

It is well known in the chaos literature that any *normally-insignificant* error in $x_0$ or $r$ in this relation in the chaotic region (e.g., when $x_0 = 0.5$, and $r = 3.97$) rapidly grows, leading to arbitrary large error in $x$ prediction (remembering that, $0 < x_n < 1$). In other words, the subject's impotence on accurate specification of ($x_0$, $r$) results in unpredictability of this chaotic physical reality[30]. Now, let us assume that ($x_0$, $r$) has been exactly specified as (0.5, 3.97), and further analyse the response of Equation (1). Figure 1 depicts the absolute difference in $x_n$ estimation employing various numbers of decimal digits compared to the estimation using 300 decimal digits (300 D.D.). The figure shows that if, for example, 200 D.D. are considered in the analysis, the difference would be negligible (smaller than 1e-15) even after 500 iterations ($n = 500$). While similar condition holds for 100 D.D. up to 364 iterations, the difference starts rapidly increasing, and reaches to more than 0.1 after 419 iterations (14 orders of magnitude increase in 55 iterations!). The figure clearly shows that the smaller the number of decimal digits considered, the sooner the emergence of drastic error.

---

[30] One can question the accuracy of such chaotic models in representing the corresponding physical realities (the notion of model faithfulness in Bishop [2017]). If it is recalled, however, that chaos studies emerged from analysing complex physical systems (Cencini et al. 2021, Ch. 7), and many chaotic characteristics have been reported in natural phenomena (Bishop 2017), the credibility of chaotic models, at least based on our current understanding of this phenomenon, is justified.

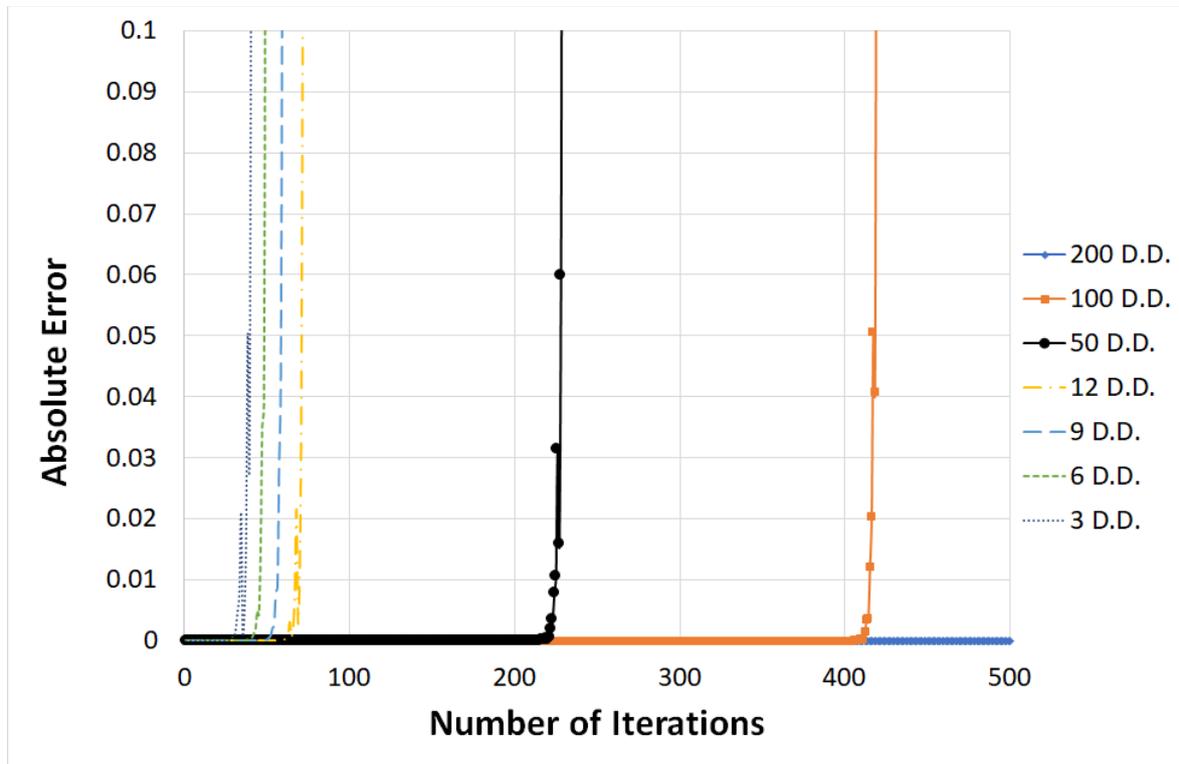

*Figure 1*. The absolute difference between $x_n$ estimated from Equation (1) using various numbers of decimal digits with estimation using 300 decimal digits ($x_0 = 0.5$, $r = 3.97$)

One could argue that it only presents an "epistemic form of nondeterminism" for such chaotic systems (Bishop 2017)[31]; however, a subtler issue is at stake here: how many significant digits are in action in the nature?[32] How many significant digits of the nature can we perceive? Nine significant digits are well above the accuracy level normally employed in engineering design practices, while most fundamental physical constants (NIST 2019) have only been determined with less than 12 significant digits[33]. The drastic error in response estimation emerges in these cases well before 100 iterations (Figure 1). As the extreme case, let us consider geometrical evaluation of Equation (1) by stretching [0,1] domain to the size of the observable universe ($O(1e+27)$ m) and performing the analysis with the accuracy of the size of smallest known particles (quarks, $O(1e-19)$ m). Figure 1 depicts that even in this extremely over-imaginative hypothesis, the drastic error in estimating the response emerges in less than 230 iterations[34]! Interestingly, an analytical solution (Equations [2]) can also be presented for this system, and it is of no help, suggesting that even finding the analytical solutions of chaotic systems (turbulence included[35]) will not bring them closer to determinism. A chaotic system is ontically indeterministic.

---

[31] Since any epistemic identification has basically a limited accuracy and many subordinate parameters are normally dropped during the analysis, the chaotic systems have epistemic indeterminacy as well.

[32] According to our deterministic presumptions, an "accurate" equation, e.g., Newton's laws of motion, is "unlimitedly" accurate. This is, of course, merely an ontological assumption within a positivistic worldview/paradigm, which could be questioned for a natural reality. It is, at least, known that in the quantum level or in the realm of relativity physics, defending such accuracies becomes a challenging task.

[33] These can still be considered as indicators of an epistemic nondeterminism, arguing that Laplace's demon, knowing everything with indefinite accuracy, could accurately predict the outcome of this chaotic system after arbitrary large iterations. I will discuss more on this metaphysically-oriented assumption in Section 6.

**4.3 Type-II examples are epistemically indeterministic.**

The indeterminacy of Type-II examples is more counter-intuitive. The permeability of a porous structure, as an example from this type, can be accurately measured, and it basically stays invariant in time, at least theoretically[36]. The indeterminacy, however, emerges in epistemic level for Type-II phenomena. The physical reality in these cases, as discussed in Sections 2.3 and 2.4, cannot be accurately *translated* through a limited number of parameters, after many decades of active research. In many cases, at least for those discussed in the mentioned sections, it is normally a trivial task to construct a physical reality (or a computer model of it) having the same exact values of the parameters developed for identification, yet significantly different overall response[37]. The response of the phenomenon in the ontic level is unique, but it is non-unique in the epistemic level, because the relation(s) between ontic and epistemic identifications is not accurate enough for such physical realities. In other words, ontically-different physical realities are perceived epistemically indistinguishable by the subject. The experimenter/subject can only see a blurred image of the physical reality in the epistemic level. As will be discussed in the next section, the possibility of this type of challenge to determinism has even been the concern of Leibniz.

**4.4 Indeterministic phenomena are not limited to the presented samples.**

I speculate that indeterminism, specifically indeterminacy Type II, could be found in many more examples in the nature. I did not find Type-II examples presented above while studying determinism in the philosophy of science. It emerged through an inverse process; the inherent impossibility of achieving an accurate estimation of the permeability of a porous medium, gradually surfaced during my research, was, in fact, my gateway into the determinism literature. It is reasonable to expect, therefore, that it is not the only entry point to this field.

---

[34] If Popper's interpretation of the role of geometry for Plato (Popper 1966, Addenda) is adopted, the instability of chaotic systems against truncation errors can also lead to serious flaws in Plato's project of geometrisation of the cosmology since even geometrical language is not accurate enough for a chaotic system.

[35] The decisive role of significant digits in modelling chaos also questions the possibility of reproducing an exact physical turbulence through direct numerical simulations of turbulence. More importantly, it fundamentally questions the Millennium Prize for analysing Navier-Stokes equations. Let's suppose even an analytical solution for Navier-Stokes equations is proposed. The turbulence still remains indeterministic, much like the analytical solution of chaotic Equation (1), presented in Equations (2), which cannot result in a deterministic answer (one output for one input).

[36] In practice, as one can guess, reaching to such invariancy is extremely difficult since any type of chemical/physical/thermal/mechanical interaction between the fluid and the solid structure, including poroelastic effects, corrosion, or clogging, can significantly change the permeability of a porous medium, as is mostly the case in the natural or designed systems.

[37] As an example, for fluid transport in porous media, it is trivial to reconstruct a porous structure with a given porosity, tortuosity, particle size, pore surface area, and presumably any other defined factor, yet with significantly different (even orders-of-magnitude larger or smaller) permeability values; see Rezaei Niya et al. (2021).

While a detailed review of the related literature for any specific physical reality is required prior to any conclusion, it is conjectured that related or similar engineering problems could also bring some obstacles in front of determinism: e.g., a) friction, in essence, results from relative translation of two uneven surfaces, which are not much dissimilar to fracture surfaces. It is, therefore, reasonable to expect to only reach to a limited accuracy in response predictions, specifically for static friction; b) the mechanical behaviours of materials, such as elastic/plastic deformation, fatigue, and buckling have long been known to show statistical characteristics, and have been treated as such in the literature (Budynas and Nisbett 2006); and c) in essence, any phenomenon analysed by employing the techniques developed to manage the indeterminacy of the nature, discussed in the next section, has potentially some indeterministic aspect.

## 5. Indeterminacy in engineering design process

Indeterminacy of the nature has long been accepted in design process, and a well-established literature has been developed to deal with the resulted ambiguities. I assume it is by now clear for the readership of this text that by indeterminacy of a physical reality, the statistical nature of the phenomenon is intended. In other words, while in the long run, a reasonably accurate estimation of the average response of the phenomenon can be obtained in most cases, the accurate prediction of a specific response is, at least for the samples discussed, out of question[38]. The determinability of the physical reality, it seems, varies in a spectrum ranging from a strict Laplacian predictability to a pure stochastic coin-tossing-like state, depending on the specific characteristics of each physical reality, depth of prediction aimed for, and certainly sensitivity of the process to perturbations in initial and boundary conditions.

Uncertainty in engineering design analyses has always been considered as "abound". Aside from uncertainties arisen from the nature of a design process such as misuse or uncontrolled manufacturing accidents with locally-concentrated effects, the natural indeterminacy of the related physical realities has also been regulated, normally with employing slightly different phrases, such as uncertainty in "composition of material and the effect of variation on properties", "validity of mathematical models used to represented reality", and "uncertainty as to the length of any list of uncertainties" (Budynas and Nisbett 2006, 22). Parameters such as "design factor", "factor of safety", or "safety factor" have been considered in the analysis to collectively represent uncertainties in any engineering design. In some design conditions, specifically when a life-threatening danger is probable, safety factors as high as 12 have been recommended (p. 898). Similar concerns about unpredictability have been dealt with employing feedback loops in control systems to control the effects of "unpredictable disturbances" (Ogata 2002, 3).

---

[38] Clearly, even for the examples provided, accurate prediction is not always impossible. Figure 1 depicts that even the chaotic system of Equation (1) can be accurately predicted up to 20 iterations even with 3 significant digits.

# 6. Historical and philosophical dimensions of determinism

Determinism has a long standing in the history of philosophy. Determinism, or at least some narration of it, can be found in ancient Greek, in Leucippus's fragments, Timaeus of Plato, and quotes remained from Zeno of Citium (see Cencini et al. 2021, Ch. 10). It was after enlightenment, however, that determinism, along with scientism, elevated from *an episteme* among or in parallel with other ways of knowing, to *the Ousia*, the only valid, credible, and legitimate description of the real world. More precisely, it was only in the last two centuries or so that determinism has established such a central rule in human understanding of the world; the first generation of the enlightened scientists, prominent figures such as Newton and Leibniz included, were theologians as well, and their scientific endeavours were not necessarily philosophically independent from their religious speculations, as will be discussed in the following. The heavy shadow of God has always been present in the disputes around determinism, interestingly on both sides, once against and once in defense of determinism, in different periods of time. During the last century or so, determinism, at least in the realm of popular science, has been considered to stand against any metaphysical interpretation, starting with the attack to free will (e.g., Ruelle 1991, Ch. 5). Determinism, on the other hand, if interpreted as a secularised and naturalistic narration of predestination[39], could be justified as a continuation of the Protestant tradition. Interestingly, the most well-known proponents of determinism have been emerged from Anglo-Saxon culture and French positivist tradition, both heavily influenced by strict Calvinist predestination beliefs for centuries (Merton 1938; Weber 2001).

The philosophical literature of determinism is not fundamentally unfamiliar with the challenges proposed in this article. Wherever the chaotic systems are introduced, their "deterministic, but unpredictable" characteristic has been emphasised (Bishop 2017; Cencini et al. 2021; Moritz 1995; Ruelle 1991; Strogatz 1994). This apologia for unpredictability of chaotic systems has been, in fact, developed by no one other than Henri Poincaré himself. A true *philosophe* with strong social and political interests (Cencini et al. 2021, Ch. 22), Poincaré in his widely acclaimed book, *Science and Method*, clarifies that he leaves no space for stochasticity, interprets determinism as possibility of unlimited accuracy in prediction at least in the inorganic world, and believes that any unpredictability is because of our approximate estimation of the initial conditions (Poincaré 1952):

> "The ancients distinguished between the phenomena which seemed to obey harmonious laws, established once for all, and those that they attributed to chance… In each domain the precise laws did not decide everything, they only marked the limits within which chance was allowed to move…
>
> But this conception is not ours. *We have become complete determinists, and even those who wish to reserve the right of human free will at least allow determinism to reign undisputed in the inorganic world*. Every phenomenon, however, trifling it be, has a cause, and a mind infinitely powerful and infinitely well-informed concerning the laws of nature could have foreseen it from the beginning of the ages. If a being

---

[39] This interpretation emphasises the inherent potentialities of determinism to be considered as an ontology, even for pre-enlightenment era.

> with such a mind existed, we could play no game of chance with him; we should always lose.
>
> For him, in fact, the word chance would have no meaning, or rather there would be no such thing as chance. That there is for us is only an account of our frailty and our ignorance. And even without going beyond our frail humanity, what is chance for the ignorant is no longer chance for the learned. *Chance is only the measure of our ignorance*"[40] (pp. 64-65).
>
> "If we knew exactly the laws of nature and the situation of the universe at the initial moment, we could predict exactly the situation of that same universe at a succeeding moment. But, even if it were the case that the natural laws had no longer any secret for us, we could still only know the initial situation *approximately*… it may happen that small differences in the initial conditions produce very great ones in the final phenomena… Prediction becomes impossible, and we have the fortuitous phenomenon" (p. 68).

The current interpretation of determinism is not significantly different from Poincaré's ideas, quoted here (see, as an example, Bishop [2017]). As a "complete determinist", Poincaré does not accept any ontic indeterminacy (Type I) or any ontic limit for determinism. An inorganic world, according to Poincaré, if far enough from human's free will, follows exactly what has been predestined for it. Unless some type of "free will" is assumed for any type of organism, Poincaré's determinism questions Darwinian evolution as no random phenomenon is possible. An "infinitely powerful" and "well-informed" mind "could have foreseen" evolution, at least up to the rise of "free will", "from the beginning of the ages". From there, then, one should either believe that free will is only an illusion and deduce an absolute predestination belief, or consider the dawn of free will as an unnatural, even a supernatural, phenomenon, which is again against Darwinism and also Poincaré's natural causation[41].

The masked metaphysical belief in the "deterministic, but unpredictable" phrase also appears between the lines of Poincaré's writing. If a physical reality is unpredictable for the experimenter, then for whom is it deterministic? Poincaré's need for a "a mind infinitely powerful and infinitely well-informed" to conquer "our frailty and our ignorance" depicts that determinism, according to him, is only possible with a direct aid of God or one of his philosophical images, Nietzsche's Übermench or Laplace's demon. Poincaré, in fact, is heavily influenced by Laplace in his interpretation of determinism. It was Laplace who stated that "… the word chance expresses only our ignorance on the causes of phenomena" (Cencini et al. 2021, 126), and for an all-knowing intelligence "nothing would be irregular" (p. 127). It is surprising that this need for a superhuman for determinism has not been questioned by philosophical wing of positivism. Logical positivists have considered any metaphysical concept on the unreachable boundary of the world. According to Wittgenstein (1922), "[m]ost propositions and questions, that have been written about philosophical matters, are not false, but senseless." (4.003)[42], and "[t]he right method of philosophy would be this. To say nothing except what can be said, *i.e.* the propositions of natural science, *i.e.* something that has nothing to do with philosophy…" (6.53). He specifically clarifies that "[t]he proposition in which there is mention of a complex, if this does not exist, becomes not nonsense but simply false" (3.24). The Poincaré's and Laplace's determinism, therefore, because of its need to a complex that does not exist, is "simply false", according to positivist philosophers!

---

[40] Italics are mine here.

[41] Drawing on a philosophical interpretation of the history of industrialisation, Veblen (1904, Ch. 9) has also discussed the metamorphosis of the concept of causality between Darwinism and prior scientific literature.

Acknowledging "our frail humanity" by Poincaré revitalises two fundamental questions facing determinism, once had been responded by Leibniz, employing his theological approach, through his well-known principles of sufficient reason and identity of indiscernibles: 1.a) How can one make sure that the nature, specifically the yet-unknown side of the nature, follows causality and reasoning? More importantly, 1.b) how can one make sure that it follows *our understanding* of causality and *our* reasoning? And, 2) how can the experimenter make sure that they can distinguish all ontically distinct phenomena? In other words, are all ontically different phenomena empirically distinguishable[43]? Many primitive indications suggest that these questions cannot be easily overlooked: a) human being is only one entity in diverse spectrum of being in the nature; b) human's senses are significantly less powerful than many other animals; and c) human progress has only been along its ontological understanding of the world[44] by invigorating its dominant senses. It is just enough to compare human progress in reinforcing its vision, an important sense for a human, to its smelling capabilities, a clearly less salient sense for humans. Despite such evident indications, the main line of reasoning to defend determinism against above-mentioned questions or similar ones is still what has been elaborated more than three centuries ago by Leibniz, based on his metaphysical assumptions (Ballard 1960; Maudlin 1993; Solari and Natiello 2019). This inherent assumption of human's place being above, beyond, and in charge of the nature, still fervently defended either explicitly or implicitly, seems to be another historical heritage of positivism from Abrahamic religious traditions.

Leibniz's responses to the above-mentioned questions stand, by any definition, outside of positivism. His principle of sufficient reason, asserting that "there must be a sufficient reason why everything is as it is and not otherwise" (Ballad 1960, 52), and principle of identity of indiscernibles, declaring that "there cannot exist two indistinguishable entities in nature" (p. 53), are traditionally considered to be developed in Leibniz-Clarke correspondence in 1715-1716 (Ballard 1960; Maudlin 1993; Solari and Natiello 2019). The outlines of both principles, however, can be traced back in Leibniz's *Discours de métaphysique* (1686). Leibniz's ideas can be summarised as follows:

---

[42] Interestingly, similar stance has been taken by Hobbs, close to three centuries earlier: "…if a man should talk to me of a round quadrangle…, I should not say he were in error; but that his words were without meaning; that is to say, absurd" (Hobbs 1651, 28).

[43] As I discussed before, Type-II examples seem to be epistemically indistinguishable, while ontically different, at least based on our current understanding of these phenomena.

[44] To be accurate, some recently-developed technologies, such as RADAR and active SONAR, are beyond extending human's senses. These systems are, however, quite recent in human's history of understanding, and are normally employed by very limited users. It should also be noted that, at least these examples, are again ontically advancing human vision, even if epistemically employing a different approach.

- Human being has an indisputable place in the universe: "*... nous croyons que Dieu n'a fait le monde que pour nous, c'est un grand abus, quoiqu'il soit très véritable qu'il l'a fait tout entier pour nous, et qu'il n'y a rien dans l'univers qui ne nous touche et qui ne s'accommode aussi aux égards qu'il a pour nous...*"[45] (19).

- God is infinitely wise, and his actions are in supreme perfection: "*...il s'ensuit que Dieu possédant la sagesse suprême et infinie agit de la manière la plus parfait...*"[46] (1) and "*Dieu agit toujours de la manière la plus parfaite et la plus souhaitable qu'il soit possible*"[47] (4).

- Neither God nor humans can do anything irregular: "*Dieu ne fait rien hors d'ordre... Ce qui est si vrai que, non seulement rien n'arrive dans le monde qui soit absolument irrégulier, mais on ne saurait même rien feindre de tel... Ainsi on peut dire que, de quelque manière que Dieu aurait créé le monde, il aurait toujours été régulier et dans un certain ordre général*"[48] (6).

- As an extension of what St. Thomas had said before about angels and intelligences, two indistinguishable, yet ontically-different substances cannot exist as well: "*… il n'est pas vrai que deux substances se ressemblent entièrement et soient différentes solo numero, et que ce que saint Thomas assure sur ce point des anges ou intelligences (quod ibi omne individuum sit species infima) est vrai de toutes les substances…*"[49] (9).

The above discussions only reiterate the fact that determinism, and positivism in general, is heavily indebted to the metaphysical beliefs and Euro-centric Judeo-Christian traditions through which it has been cultivated. Positivism has, undoubtedly, facilitated the progress of human's knowledge and understanding of the nature during the last centuries. However, the ingenious internal harmony of positivistic perspective has, sometimes, led its practitioners to overlook the "-ism" side of its identity, and to publicise positivism against, rather than along, other epistemes. The social, political, and cultural conflicts, resulted from modernity, could have potentially been managed in calmer and more dialogical environments if deep-rooted Christianity in positivism has been publicly considered. Humanity, unfortunately, will soon pay the harshest price for such misunderstanding, in not centuries to come, but only the following handful of decades, as the consensual denial of climate disaster by the public rests heavily on now-internalised general misbelief in superiority of humans to the nature, predictability of the nature's future, and absolute authority of deterministic science over the inexperienced conditions for human's life on earth.

---

[45] "…it is a great mistake to think that God made the world only for us, although it is true that he did make it—all of it—for us, and that there is nothing in the universe that does not touch us, and which is not also adjusted to fit the concern he has for us…" All translations are from Jonathan Bennett's edition (2017).

[46] "…the actions of God, who is supremely—indeed infinitely—wise, are completely perfect."

[47] "… God always acts in the most perfect and desirable way possible."

[48] "… God does nothing that isn't orderly… Indeed, not only does nothing absolutely irregular ever happen in the world, but we cannot even feign such a thing… So one can say that no matter how God had created the world, it would have been regular and in some general order."

[49] "… it is never true that two substances are entirely alike, differing only in being two rather than one, [and that what St. Thomas says on this point regarding angels and intelligences (quod ibi omne individuum sit species infima) is true of all substances…]". The sentence in the bracket has been dropped in Bennett's translation; the translation from Wikisource has been quoted for this sentence. To be accurate, Leibniz here emphasises the impossibility of the existence of two ontically-identical substances, without stating the impossibility of the existence of two indiscernible, but ontically-different substances (the subtle difference lies in the incompetency of the subject). The latter has been discussed more clearly in Leibniz-Clarke correspondence (Ariew 2000). However, when the former statement is combined with the superior place of human being in the universe in Leibniz's eyes ("*nous croyons que Dieu n'a fait le monde que pour nous*"), the latter can be easily expected.

## 7. Beyond positivism in Newtonian mechanics

Challenging positivism in Social Sciences and Humanities is prevalent nowadays in the literature (O'leary 2004; Phillips and Burbules 2000; Reichardt and Rallis 1994; Robson and McCartan 2016), and diverse epistemological and ontological assumptions have been developed and are employed in these fields. The path beyond positivism, normally, passes through post-positivism, which, in fact, includes any approach questioning positivistic assumptions (O'leary 2004, 6; Phillips and Burbules 2000, Ch. 2; Robson and McCartan 2016, 22). Post-positivism believes that the reality "can only be known imperfectly and probabilistically in part because of the researcher's limitations" (Robson and McCartan 2016, 22)[50]. The discussions provided in this text, therefore, could be considered a step beyond positivism towards post-positivistic ontology. It should be interpreted as an invitation to re-examining the hidden positivistic assumptions, specifically in engineering applications. The following recommendations, drawing on this perspective, are proposed here:

1. The accuracy limit of presented relations, constants, and equations, specifically those directly related to the physical reality, needs to be explored and reported. Ideally, the estimated levels of accuracy possible in different applications could be reported: e.g., a) the expected accuracy level in a general laboratory by a lay person; b) the expected industrial accuracy level; c) the accuracy level possible in a well-controlled environment by an experienced experimenter; and d) the theoretical maximum possible accuracy. The last, which is the most related here, needs a thorough theoretical-modeling-experimental analysis. While some basically-similar estimations are already employed in design processes through error analysis, the inherent indeterminacy level of physical realities (level d), such as Type-II examples, is unclear in many engineering applications. Acknowledging the indeterminacy of such systems shifts the research questions from developing new relations in the hope of achieving higher accuracies (e.g., for permeability estimation of a general porous structure), towards estimating the statistical characteristics of target parameters, e.g., average value, standard deviation, skewness, kurtosis, modality, and the probability of outliers.

2. Any attempt towards developing an accurate universal relation for a wide range of physical realities (e.g., all types of fractures, rocks, porous structures) is based on the inherent assumption of existence of such physical harmony, which is, in essence, more inclined towards the "-ism" side of positivism. While certainty of impossibility of such general relations in all different applications without any investigation again originates from ideological assumptions, prior to any thorough research on *discovering* one-tool-for-all relations, the possibility of their existence, according to the available experimental results, needs to be studied.

---

[50] As one can guess, a slight difference between the definitions of post-positivism provided in various references can be expected. O'leary's (2004) definition, emphasising that "the world may not be 'knowable'" and it is "open to interpretation" (p. 6), is closer to interpretivism than what is meant by post-positivism here.

3. Similarly, the assumption of applicability of one abstract mathematical concept (e.g., fractal theory) or, in general, a specific type of a deterministic correlation, to every complex physical reality is mainly based on ideological side of positivism[51]. A thorough analysis of at least possibility of the emergence of similar characteristics to those assumed in such mathematical concepts (e.g., self-similarity in fractal theory) during the natural formation or manufacturing process of the specific physical reality under investigation needs to be prioritised in such analysis.

4. Positivism, as an ideological-historical-philosophical and pragmatic framework, needs to be introduced to the natural sciences and engineering researchers, at least in the graduate level[52]. It should be clarified that fundamental research assumptions, like possibility of accurate prediction of an unknown phenomenon, are not well rooted in any *exact* mathematical/theoretical analysis, experimental studies, or even fundamental laws of physics. Two research directions, along the following questions, could be pursued from here: 1) What if such assumptions are incorrect or inaccurate? What experimental or theoretical evidences support such claims? What adjustments are required in our description of physical realities? And, 2) could any philosophical interpretation be developed *from within positivism* to re-establish the metaphysics-based assumptions, essentially invented by Leibniz, employed in positivism?

5. STEM (Science, Technology, Engineering, and Mathematics) education is, in essence, based on assuming positivism being *the Ousia*. Including a wider coverage of philosophical, social, and historical background of the topics in curricula could help the practitioners to explore physical reality with less solid and more translucent preconceptions.

---

[51] The limitations of fractal dimension concept for characterising physical fractures have already been extensively discussed in the literature (e.g., Ge et al. 2014; Xie et al. 1997).

[52] I have not encountered such a course during my residency in engineering departments at Stanford University in the US, McGill University, University of British Columbia, University of Manitoba, and Concordia University in Canada, RMIT University in Australia, and Sharif University of Technology and Tehran Polytechnic in Iran.

**References**

Aghdam, M. M., Farahani, M. R. N., Dashty, M., and Rezaei Niya, S. M. 2006. Application of generalized differential quadrature method to the bending of thick laminated plates with various boundary conditions. In *Applied Mechanics and Materials* (Vol. 5, pp. 407-414). Trans Tech Publications Ltd.

Alam, M. A. 1978. Critique of positivism in the natural sciences. *Social Scientist*, 67-78.

Alfonsi, G. 2011. On direct numerical simulation of turbulent flows. *Applied Mechanics Reviews*, *64*(2).

Ariew, R. (Ed.) 2000. *G. W. Leibniz and Samuel Clarke correspondence*. Hackett Publishing Company.

Ballard, K. E. 1960. Leibniz's theory of space and time. *Journal of the History of Ideas*, *21*(1), 49-65.

Bear, J. 1972. *Dynamics of fluids in porous media*. Dover Publications.

Bear, J., and Cheng, A. H. D. 2010. *Modeling groundwater flow and contaminant transport* (Vol. 23, p. 834). Dordrecht: Springer.

Bear, J., Tsang, C. F., and De Marsily, G. 1993. *Flow and contaminant transport in fractured rock*. Academic Press.

Bishop, R. 2017. "Chaos", *The Stanford Encyclopedia of Philosophy*, Zalta, E. N. (ed.), URL = <https://plato.stanford.edu/archives/spr2017/entries/chaos/>.

Bloor, D. 2004. Sociology of scientific knowledge. In *Handbook of epistemology* (pp. 919-962). Springer, Dordrecht.

Brown, G. O. 2002. Henry Darcy and the making of a law. *Water Resources Research*, *38*(7), 11-1.

Budynas, R. G., and Nisbett, J. K. 2006. *Shigley's mechanical engineering design*. McGraw-Hill.

Cencini, M., Puglisi, A., Vergni, D., and Vulpiani, A. 2021. *A random walk in physics*. Springer.

Colagrossi, A., Marrone, S., Colagrossi, P., and Le Touzé, D. 2021. Da Vinci's observation of turbulence: A French-Italian study aiming at numerically reproducing the physics behind one of his drawings, 500 years later. *Physics of Fluids*, *33*(11), 115122.

Collins, H. M., and Pinch, T. 1998. *The Golem: What you should know about science*. Cambridge University Press.

Creath, R. 2023. "Logical Empiricism", *The Stanford Encyclopedia of Philosophy*, Zalta, E. N., Nodelman, U. (eds.), URL = <https://plato.stanford.edu/archives/win2023/entries/logical-empiricism/>.

Dryden, H. L. 1939. Turbulence and diffusion. *Industrial & Engineering Chemistry*, *31*(4), 416-425.

Earman, J. 1986. *A primer on determinism*. D. Reidel Publishing Company.

Frigg, R., and Hartmann, S. 2020 "Models in Science", *The Stanford Encyclopedia of Philosophy*, Zalta, E. N. (ed.), URL = <https://plato.stanford.edu/archives/spr2020/entries/models-science/>.

Frigg, R., and Nguyen, J. 2021. "Scientific Representation", *The Stanford Encyclopedia of Philosophy*, Zalta, E. N. (ed.), URL = <https://plato.stanford.edu/archives/win2021/entries/scientific-representation/>.

Ge, Y., Kulatilake, P. H., Tang, H., and Xiong, C. 2014. Investigation of natural rock joint roughness. *Computers and Geotechnics*, *55*, 290-305.

Gleick, J. 1987. *Chaos: Making a new science*. Penguin.

Hadamard, J. 1902. Sur les problèmes aux dérivées partielles et leur signification physique. *Princeton university bulletin*, 49-52.

Hadamard, J. 1915. *Four lectures on mathematics* (No. 5). Columbia University Press.

Hadamard, J. 1923. *Lectures on Cauchy's problem in linear partial differential equations*. Yale University Press.

Hinze, J. O. 1959. *Turbulence*. McGraw-Hill.

Hoare, F. R. 1932. Indeterminacy and indeterminism: With a suggestion for interpreting the former. *Philosophy*, *7*, 394-403.

Hobbs, T. 1651. *Leviathan*. London.

Holmes, P., Lumley, J. L., Berkooz, G., and Rowley, C. W. 2012. *Turbulence, coherent structures, dynamical systems and symmetry*. Cambridge university press.

Hunt, J. C. R., Sandham, N. D., Vassilicos, J. C., Launder, B. E., Monkewitz, P. A., and Hewitt, G. F. 2001. Developments in turbulence research: a review based on the 1999 Programme of the Isaac Newton Institute, Cambridge. *Journal of Fluid Mechanics*, *436*, 353-391.

Ichikawa, Y., and Selvadurai, A. P. 2012. *Transport phenomena in porous media: Aspects of micro/macro behaviour*. Springer Science & Business Media.

Jacoby, R. 1987. *The last intellectuals: American culture in the age of academe*. Basic Books.

Kellert, S. 1993. *In the wake of chaos*. Chicago University Press.

Korolev, A. 2007. Indeterminism, asymptotic reasoning, and time irreversibility in classical physics. *Philosophy of Science*, *74*, 943-956.

Kuhn, T. S. 1962. *The structure of scientific revolutions*. University of Chicago Press: Chicago.

Lindelöf, E. 1894. Sur l'application de la méthode des approximations successives aux équations différentielles ordinaires du premier ordre. *Comptes rendus hebdomadaires des séances de l'Académie des sciences*, *116*, 454-457.

Malament, D. B. 2008. Norton's slippery slope. *Philosophy of Science*, *75*, 799-816.

Maudlin, T. 1993. Buckets of water and waves of space: Why spacetime is probably a substance. *Philosophy of science*, *60*(2), 183-203.

McMullin, E. 1985. Galilean idealization. *Studies in History and Philosophy of Science Part A*, *16*(3), 247-273.

Menke, H. P., Maes, J., and Geiger, S. 2021. Upscaling the porosity–permeability relationship of a microporous carbonate for Darcy-scale flow with machine learning. *Scientific Reports*, *11*(1), 1-10.

Merton, R. K. 1938. Science, technology and society in seventeenth century England. *Osiris*, *4*, 360-632.

Mill, J. S. 1973. *Auguste Comte and Positivism*. University of Michigan Press.

Moritz, H. 1995. *Science, mind and the universe, an introduction to natural philosophy*. Wichmann.

Nield, D. A., and Bejan, A. 2006. *Convection in porous media* (Vol. 3). New York: springer.

NIST 2019 National Institute of Standards and Technology, U.S. Department of Commerce, URL = < https://physics.nist.gov/cuu/Constants/>.

Norouzi, M., Rezaei Niya, S. M., Kayhani, M. H., Shariati, M., Karimi Demneh, M., and Naghavi, M. S. 2012. Exact solution of unsteady conductive heat transfer in cylindrical composite laminates. *Journal of heat transfer*, *134*(10).

Norton, J. 2003. Causation as folk science. *Philosopher's imprint*, *3*, 1-22.

Norton, J. 2008. The dome: An unexpectedly simple failure of determinism. *Philosophy of Science*, *75*, 786-798.

Ogata, K. 2002. *Modern control engineering*. Pearson.

O'leary, Z. 2004. *The essential guide to doing research*. Sage.

Petrovitch, C. L., Nolte, D. D., and Pyrak-Nolte, L. J. 2013. Scaling of fluid flow versus fracture stiffness. *Geophysical Research Letters*, *40*(10), 2076-2080.

Petrovskii, I. G. 1966. *Ordinary differential equations*. Prentice-Hall.

Phillips, D. C., and Burbules, N. C. 2000. *Postpositivism and educational research*. Rowman & Littlefield.

Poincaré, H. 1952. *Science and method*. Thomas Nelson and Sons.

Polanyi, M. 1962. *Personal knowledge*. Routledge.

Popper, K. R. 1966. *The open society and its enemies*. Princeton University Press.

Pyrak-Nolte, L. J., and Morris, J. P. 2000. Single fractures under normal stress: The relation between fracture specific stiffness and fluid flow. *International Journal of Rock Mechanics and Mining Sciences*, *37*(1-2), 245-262.

Pyrak-Nolte, L. J., and Nolte, D. D. 2016. Approaching a universal scaling relationship between fracture stiffness and fluid flow. *Nature communications*, *7*(1), 1-6.

Reichardt, C. S., and Rallis, S. F. 1994. The relationship between the qualitative and quantitative research traditions. *New Directions for Program Evaluation*, *1994*(61), 5-11.

Reynolds, O. 1883. XXIX. An experimental investigation of the circumstances which determine whether the motion of water shall be direct or sinuous, and of the law of resistance in parallel channels. *Philosophical Transactions of the Royal society of London*, (174), 935-982.

Reynolds, O. 1894. IV. On the dynamical theory of incompressible viscous fluids and the determination of the criterion. *Cambridge Phil. Trans*, 123-164.

Rezaei Niya, S. M., and Andrews, J. 2022. On charge distribution and storage in porous conductive carbon structure. *Electrochimica Acta*, *402*, 139534.

Rezaei Niya, S. M., Hejabi, M., and Gobal, F. 2010. Estimation of the kinetic parameters of processes at the negative plate of lead-acid batteries by impedance studies. *Journal of Power Sources*, *195*(17), 5789-5793.

Rezaei Niya, S. M., and Hoorfar, M. 2014. Process modeling of the ohmic loss in proton exchange membrane fuel cells. *Electrochimica Acta*, *120*, 193-203.

Rezaei Niya, S. M., and Hoorfar, M. 2015. Process modeling of electrodes in proton exchange membrane fuel cells. *Journal of Electroanalytical Chemistry*, *747*, 112-122.

Rezaei Niya, S. M., and Hoorfar, M. 2016. On a possible physical origin of the constant phase element. *Electrochimica Acta*, *188*, 98-102.

Rezaei Niya, S. M., Naghshbandi, S., and Selvadurai, A. P. S. 2021. On non-uniqueness issues related to the permeability of a porous medium with a random porous structure. *arXiv e-prints*, *arXiv:2112.11762*.

Rezaei Niya, S. M., Phillips, R. K., and Hoorfar, M. 2016-2. Process modeling of the impedance characteristics of proton exchange membrane fuel cells. *Electrochimica Acta*, *191*, 594-605.

Rezaei Niya, S. M., and Selvadurai, A. P. S. 2017. The estimation of permeability of a porous medium with a generalized pore structure by geometry identification. *Physics of Fluids*, *29*(3), 037101.

Rezaei Niya, S. M., and Selvadurai, A. P. S. 2018. A statistical correlation between permeability, porosity, tortuosity and conductance. *Transport in Porous Media*, *121*(3), 741-752.

Rezaei Niya, S. M., and Selvadurai, A. P. S. 2019. Correlation of joint roughness coefficient and permeability of a fracture. *International Journal of Rock Mechanics and Mining Sciences*, *113*, 150-162.

Rezaei Niya, S. M., and Selvadurai, A. P. S. 2021. Modeling the approach of non-mated rock fracture surfaces under quasi-static normal load cycles. *Rock Mechanics and Rock Engineering*, *54*(4), 1885-1896.

Robson, C., and McCartan, K. 2016. *Real world research*. Wiley Global Education.

Ruelle, D. 1991. *Chance and Chaos*. Princeton University Press.

Scheidegger, A. E. 1963. *Principles of geodynamics*. Springer-Verlag.

Scheidegger, A. E. 1974. *The physics of flow through porous media*. University of Toronto Press.

Schlichting, H., and Gersten, K. 2017. *Boundary layer theory*. Springer.

Selvadurai, A. P. S., Couture, C. B., and Rezaei Niya, S. M. 2017. Permeability of wormholes created by CO2-acidized water flow through stressed carbonate rocks. *Physics of Fluids*, *29*(9), 096604.

Selvadurai, A. P. S., and Rezaei Niya, S. M. 2020. Effective thermal conductivity of an intact heterogeneous limestone. *Journal of Rock Mechanics and Geotechnical Engineering*, *12*(4), 682-692.

Selvadurai, A. P., and Suvorov, A. P. 2017. *Thermo-poroelasticity and geomechanics*. Cambridge University Press.

Smith, M. W. 2014. Roughness in the earth sciences. *Earth-Science Reviews*, *136*, 202-225.

Smith, P. 1998. *Explaining Chaos*. Cambridge University Press.

Solari, H. G., and Natiello, M. A. 2019. A constructivist view of newton's mechanics. *Foundations of Science*, *24*(2), 307-341.

Strogatz, S. H. 1994. *Nonlinear dynamics and chaos: with applications to physics, biology, chemistry, and engineering*. CRC press.

Tabrizi, H. B., Rezaei Niya, S. M., and Fariborz, S. J. 2004. Investigation of turbulent plane mixing layer using generalized differential quadrature. *12$^{th}$ Annual Computational Fluid Dynamics Conference*, Ottawa, Canada.

Taylor, G. I. 1914. I. Eddy motion in the atmosphere. *Philosophical Transactions*, *215*(523-537), 1-26.

Taylor, G. I. 1920. Diffusion by continuous movements. *Proceedings of the London mathematical society*, *2*(1), 196-212.

Taylor, G. I. 1935. Statistical theory of turbulence. *Proceedings of the Royal Society of London. Series A-Mathematical and Physical Sciences*, *151*(873), 421-444.

Tikhonov, A. N., and Arsenin, V. I. 1977. *Solutions of Ill-posed Problems: Andrey N. Tikhonov and Vasiliy Y. Arsenin. Translation Editor Fritz John*. Wiley.

Tikhonov, A. N., Goncharsky, A. V., Stepanov, V. V., and Yagola, A. G. 1995. *Numerical methods for the solution of ill-posed problems* (Vol. 328). Springer Science & Business Media.

Turner, J. H. 2001. " Positivism: Sociological", *International Encyclopedia of the Social & Behavioral Sciences*, Smelser, N. J., Baltes P. B. (eds.), Pergamon.

Van Strien, M. 2014. The Norton dome and the nineteenth century foundations of determinism. *Journal of General Philosophy of Science*, *45*, 167-185.

Veblen, T. 1904. *The Theory of Business Enterprise*. Charles Scribner's Sons.

Veblen, T. 1918. *The Higher Learning in America: A Memorandum on the Conduct of Universities by Business Men*.

Von Karman, T. 1937. On the statistical theory of turbulence. *Proceedings of the National Academy of Sciences*, *23*(2), 98-105.

Wang, L., and Cardenas, M. B. 2016. Development of an empirical model relating permeability and specific stiffness for rough fractures from numerical deformation experiments. *Journal of Geophysical Research: Solid Earth*, *121*(7), 4977-4989.

Weber, M. 2001. *The protestant ethic and the spirit of capitalism*. Routledge.

White, F. M. 2009. *Fluid mechanics*. McGraw-Hill.

Wittgenstein, L. 1922. *Tractatus logico-philosopicus*. Kegan Paul, Trench, Trubner & Co.

Xiao, T., Guo, J., Yang, X., Hooman, K., and Lu, T. J. 2022. On the modelling of heat and fluid transport in fibrous porous media: Analytical fractal models for permeability and thermal conductivity. *International Journal of Thermal Sciences*, *172*, 107270.

Xie, H., Wang, J. A., and Xie, W. H. 1997. Fractal effects of surface roughness on the mechanical behavior of rock joints. *Chaos, Solitons & Fractals*, *8*(2), 221-252.

Zammito, J. H. 2004. *A nice derangement of epistemes: Post-positivism in the study of science from Quine to Latour*. University of Chicago Press.

Zinkernagel, H. 2010. Causal fundamentalism in physics. In Suarez, M., Dorato, M. and Redei, M. (Eds.), *EPSA philosophical issues in the sciences: Launch of the European philosophy of science association*. Springer.

**Appendix: Reviewing History**

The reviewing process of this manuscript, submitted to and rejected by various journals dedicated to the philosophy and history of science, is presented in this appendix. Over the last two years from the initial submission of the text on Sep. 12$^{th}$, 2022, the manuscript has been rejected by most journals without undergoing any reviewing process. Rejections have often been issued directly by the editor, employing the most general terms possible, and in some cases, offering contradictory rationales, as presented in this document.

The text has basically remained intact from one submission to the next, with occasional footnotes added based on the limited provided reviews or my further reading on related topics. As can be seen, the manuscript has, in most cases, been simply ignored by the editors, due to the reasons and conditions certainly beyond the positivistic assumptions normally associated with the peer-review process. It is, by itself, another indication that the assumption of the disinterested, double-blinded reviewing process overlooks the inherent power structure of the established paradigm, questioning of which requires recognition and respect that can only be obtained by the "insider" community. This manuscript, along with similar ones, will therefore have little chance of being read, even by the reviewers, since the editors, as guardians of the established scientific tradition, tend to protect the accepted paradigms of the field. Whether it preserves the scientific and philosophical discussions from unworthy distractions, or it constructs a scientific conservative dictatorship suppressing any eccentric perspective, the outcome will be the same.

If any contribution from contextual conditions in the pursuit of human understanding and knowledge is presumed, it should be acknowledged that the direr the political-social-economical status of an era, the farther the scientific discourse from idealistic, meritocratic, and democratic assumptions envisaged from an enlightened age. It is particularly worrying when one reflects on the rapid large-scale changes imposed by climate disaster on human lives in a very near future. Considering the proposed epistemic "solutions" for such a huge ontic shock, for which even global north-south dichotomies are meaningless, it becomes evident that the army of academe is woefully unprepared for its existential battle!

<div style="text-align: right;">S.M.R.N.</div>

<div style="text-align: right;">May 2024</div>

**European Journal for Philosophy of Science:** EPSA-D-22-00163

Submission date: 2022-09-12

**Editor:** *We apologize that it has taken us so long to get back to you. While your paper discusses a topic of interest to our journal--determinism--it does not engage with any of the central literature on determinism in the philosophy of science and is a bit too far ranging in its scope (both as far as historical context and as far as the examples from physics are concerned). While this breadth is interesting, it comes at the expense of sufficient detail and depth.*

**Editor:** *With regret, I must inform you that we have decided that your paper cannot be accepted for publication in Journal for General Philosophy of Science. The main reason is that there is a vast literature in philosophy of physics that already present your findings as arguments against traditional Newtonian mechanics. This means that we do not reject your paper because we find important problems in your arguments, but because your arguments are not original.*

**Foundations of Science:** FODA-D-23-00060

Submission date: 2023-04-04

**Editor:** *With regret, we must inform you that, based on the advice received, we have decided that your manuscript cannot be accepted for publication in Foundations of Science.*

**Reviewer #1:** *General comment to the Author(-s)*
*\* The author must rethink-improve-rework presentations, included its lay-out, as well as for making sure it best conveys the author's ideas.*
*# The paper is interesting but its historical method and foundational-presentation are very weak. This make sure some difficulties to follow author(-s) arguing/claims.*
*# It is not clear if the author(-s) is dealing with Newton's Mechanics (1687 and succ. eds.), or dealing with 19th Newtonian Mechanics, including positivism etc., or dealing with modern Newtonian Mechanics, at present day, experimental results etc. In fact, in the paper several claiming does not belong to Newton's Mechanics (Principia, 1687), maybe they belong to modern Newtonian Mechanics.*
*# A Conclusion numbered section lacks in the paper.*
*# At this stage the author(-s) should rethink-improve-rework contents and presentations, included its lay-out sections, as well as for making sure it best conveys the author's ideas.*
*# The author(-s) is friendly advised.*

**Reviewer #2:** *This paper deals with the problem of determinism in classical physics. In particular it claims the universe not to be deterministic either epistemically or ontically. The author tries, then, to establish the deterministic root in positivism.*
*This paper is badly conceived: the given examples are not convincing to prove that classical physics is not deterministic. Other more interesting and convincing ones might have been given. For example, if mechanical energy is represented by*

$$E = \frac{\dot{x}^2}{2} - \frac{x^{2\alpha}}{2}, \alpha \geq 0$$

*when $\frac{1}{2} < \alpha < 1$, you reach an indeterministic situation within ordinary Newtonian mechanics (no need to speak of chaos, of turbulence, etc.).*
*The philosophical-interpretative part is even weaker. The author claims that the root of determinism in positivism dates back to Leibniz. He reports some opinions of Leibniz (they are not well referred. There is no critical considerations, nothing) and claims that they influenced positivism. But no proof is given, no historical and conceptual link between Leibniz and positivism is shown. The origin of positivism's determinism could be Leibniz as other authors as no author in particular, but the development itself of physics. The final section is obscure, convolute and adds nothing precise. Conclusions are completely missing. This paper has to be rejected.*

**Authors' Reply to the Editors:** I recently received the decision letter regarding my submitted manuscript FODA-D-23-00060 "Positivism in Newtonian Mechanics: The Ousia or a Historical Liability?". After two full months of reviewing process for a journal with median decision time of 13 days, I have got two review letters, which are both, in my opinion, unqualified to evaluate this work:

- Reviewer #1 has suggested the rejection of my 12,000-words manuscript with a barely-intelligible 124-words letter, full of grammatical errors, as he believes "the author(-s) should rethink-improve-rework contents and presentations, included its lay-out sections". As other reasons for his/her decision, the reviewer mentions "A Conclusion numbered section lacks in the paper." and "The author(-s) is friendly advised."! S/he adds "This make sure some difficulties to follow author(-s) arguing/claims."! The only point mentioned by Reviewer #1 is that he has questioned that the experimental results discussed and analysed in my manuscript "does not belong to Newton's Mechanics (Principia, 1687)"!, as if all the theoretical analyses developed based on Newtonian mechanics should have been found in Newton's treatise! It is clear that Reviewer #1 is fundamentally alien to the engineering literature by any sort, and is without any basic engineering-science background!
- To be fair, I should confess that Reviewer #2 seems to be at least familiar with the basic concepts discussed in the manuscript, albeit on the surface. S/he starts the review report by proposing one sample of many "thought" experiments, nowadays fashioned to discuss determinism in Newtonian mechanics. The reviewer seems not to be familiar with the general cases of such examples, discussed in detail in footnote 6, specifically the Lipschitz-indeterminism. The reviewer then questions the role of Leibniz in the establishment of determinism, a point that is so subtly accepted in the literature that it is practically impossible to find any philosophical text on determinism without mentioning Leibniz's principles (e.g., Ballard 1960; Maudlin 1993; Solari and Natiello 2019).

This manuscript is the outcome of my research and studies of the last decade, and I have chosen every single word of the text with utmost care. During my residency at Stanford and McGill universities in the US and Canada, I have discussed and published the engineering aspects of this manuscript with world-renowned scholars of these fields in various papers. I have also conducted an extensive study of the recent and original texts for the philosophical discussions before writing this manuscript. I, therefore, believe that this text deserves a thorough review by experienced philosophers of science and epistemologists. I would appreciate your consideration if you could provide this chance for this work to be re-reviewed and re-evaluated to receive detailed reviews.

(*No response was provided by the editors to this email*)

**International Studies in the Philosophy of Science:** 233336551

Submission date: 2023-06-05

**Editor:** *I regret to inform you that our reviewers have now considered your paper but unfortunately feel it unsuitable for publication in International Studies in the Philosophy of Science. For your information I attach the reviewer comments at the bottom of this email. I hope you will find them to be constructive and helpful. You are of course now free to submit the paper elsewhere should you choose to do so.*

**Reviewer #1:** *This manuscript proposes examples and doctrines opposed to determinism and what it calls "objectivism" in Newtonian mechanics, the two of which it takes to be commitments of positivism. It also speculates about positivism being based on Leibniz's metaphysics, which, it claims, are in conflict with "physical reality". I'm afraid that I wasn't convinced by any of the arguments or examples for these main points, and I am rather concerned about the manuscript's lack of scholarly contact with the literature pertaining to the positions it criticizes. In what follows I explain my main reasons for reaching these conclusions.*

*In the history of philosophy, "positivism" denotes a family of doctrines about the proper method for inquiry and the status of knowledge. Both the original version formulated by Comte in the 1830s and the logical positivism of some (proper part) of the Vienna Circle about a hundred years later saw that this involved a rejection of metaphysics. For Comte, this was a rejection of the search for true causes for lawlike regularities, particularly for social phenomena; for the logical positivists, this involved a distinction between analytic and synthetic statements (inherited from Kant), the former being a priori truths of logic and mathematics and the latter being wholly a posteriori truths of common experience. They had the ambition to use the newly developed tools of mathematical logic to show how the knowledge provided by empirical science can be understood wholly through combinations of these two categories. In particular, they had this ambition for the probabilistic claims of quantum mechanics, developed around the same time by scientists in close contact with the Vienna Circle.*

**{Authors' Reply:** The reviewer clearly follows the definition of positivism discussed in footnote 7**}**

*In any case, there is not a single piece of evidence that determinism, much less any other metaphysical doctrine (regardless of Leibnizian affinity), has been an assumption of any form of positivism. Indeed, because the rejection of metaphysics is the main thread connecting all the doctrines in this family together, if anything positivism would reject determinism as superstition or ultimately meaningless.*

**{Authors' Reply:** It is not clear how the reviewer reconciles using "the newly developed tools of mathematical logic" with rejection of determinism, as if mathematics is not inherently deterministic**}**

*The manuscript thus evinces no awareness about the actual commitments of positivism, its historical context, which post-dates the development of Newtonian mechanics by more than a*

*century, or why it came to be widely rejected. The "positivism" to which it eludes seems to be a version of a bogeyman found in some non-philosophical treatments of methodology, such as those cited in footnote 4, that has no substantive relation to positivism in fact.*

**{Authors' Reply:** The quote mentioned in footnote 7 has been borrowed from this respected reviewer**}**

*Similar remarks apply to the manuscript's claims about positivism's commitment to "objectivism", that there is no difference between "ontic and epistemic identifications" of a phenomenon, i.e., between the phenomenon (or tokens thereof) itself and descriptive models of the phenomenon. The only support given for this position is the assertion that the mathematician J. Hadamard was a positivist (despite not being an empirical scientist?) and some quotations from him in footnote 21. The quotations do not clearly support the position attributed to Hadamard, and no actual analysis of the quotations is provided.*

**{Authors' Reply:** The reviewer seems not to be familiar with the importance of Hadamard's well-posedness criteria in the engineering and science fields, or the role of Cauchy's problem in discussions on determinism**}**

*In any case, even if one were to establish that Hadamard held that there were no difference between "ontic and epistemic identifications" of a phenomenon, it would not imply that this is a general commitment of positivism. (The manuscript claims that the distinction between is "ontic identification" and "epistemic identification" is similar to, but distinct from this distinction between targets of models and models themselves, but did not explain what the differences were supposed to be. I was not able to see what the differences could be, which raises the question of whether coinage of novel jargon is really helpful here.)*

*But even if one were to extricate all references to positivism in the manuscript to focus on its real target, determinism, the manuscript's arguments would still be unsatisfactory. First, the manuscript introduces an opaque distinction between "Type I" and "Type II" indeterminacy that is never explained except that Type I involves "well accepted" unpredictability while Type II also involve unpredictability but only "deep in the details of the analyses". It \*seems\* as if the distinction has to do with how much the examples have been discussed, rather than with any intrinsic features of the examples themselves, in which case these are not different types of indeterminism at all but just classes which have received different levels of attention, a distinction that hardly seems to be of conceptual import.*

**{Authors' Reply:** The respected reviewer seems not to be familiar with the role of chaos in discussions on determinism.**}**

*Yet in section 4 it is argued that this different level of attention is correlated with determinism at different "levels". Section 4.2 doesn't apparently present any explicit argument that Type I, chaotic systems, are really indeterministic, but rather a series of suggestions and rhetorical questions. These focus on the true fact that variable computational truncation errors can imitate the sort of sensitivity to initial conditions typically used to characterize chaotic systems, suggesting that this is an "epistemic form of nondeterminism", then asking "how many significant digits are in action in the [sic] nature?" The suggestion is false: both nature and the model are deterministic, even though different computational models with different numbers of significant figures will produce different results. Moreover, the rhetorical question assumes a category mistake: nature doesn't have any significant digits because those are*

*features of our models, not nature. (The suggestion here commits the very mistake of conflating the epistemic and ontic identifications that the manuscript cautions against before!) None of the considerations in this section support the conclusion that there is any indeterminism in these examples, either at the level of the phenomena or at the level of the models.*

**{Authors' Reply:** The reviewer's "belief" in determinism of the nature is, in fact, similar to Poincaré's point of view. The question is, as discussed in Section 6 of the manuscript, if "different computational models with different numbers of significant figures will produce different results" than the nature, then for whom the nature is deterministic? The same misunderstanding is repeated about Type II examples below. The reviewer forgets that if "no" model can "ever" accurately describe a system, then the system should be considered indeterministic!**}**

*Section 4.3 suggests that there is indeterminism at the epistemic level for the Type II examples "because the relations(s) between ontic an epistemic identifications is not accurate enough for such physical realities. In other words, ontically-different physical realities are perceived epistemically indistinguishable by the subject." Put in simpler terms, different systems are often described by the same model. This is however not a form of indeterminism at all, but a fact about how there can be discrepancies between phenomena and how a model describes it. If the model makes a single prediction, it is deterministic, regardless of whether that prediction sometimes matches what is being modeled and sometimes not.*

**Authors' Reply to the Reviewer #1:** I have inserted my responses into the reviewer's text in **{Authors' Reply:** this**}** format.

A8

**Philosophy of Science:** PHOS-16696

**Editor:** *While this paper has a number of significant virtues, I am sorry to report that it is not well suited for publication in Philosophy of Science. The main reasons are that the line of argument of the paper is too broad and programmatic for the journal, and it does not make the right sort of contact with the recent literature in the philosophy of science.*

**Erkenntnis:** ERKE-D-23-00267

**Editor:** *Thank you for submitting your paper "Positivism in Newtonian Mechanics: The Ousia or a Historical Liability?" to Erkenntnis. We are sorry to have to tell you that the paper is unsuitable for publication in Erkenntnis. The subjects covered are by far too large, and the paper lacks a clear focus for our audience.*

**American Journal of Physics:** AJP23-AR-00464

**Editor:** *We have reviewed your submission, "Positivism in Newtonian Mechanics: The Ousia or a Historical Liability?," (our manuscript #AJP23-AR-00464) and determined that it is not appropriate for publication in the American Journal of Physics. Please refer to the "Information for Contributors" and the "Statement of Editorial Policy" at the AJP home page at https://www.aapt.org/Publications/AJP.*

*In particular, the guidelines state, "Manuscripts announcing new theoretical or experimental physics research results, or questioning well-established and successful theories, are not acceptable, since AJP is a pedagogical journal. Such manuscripts should be submitted to a research journal for consideration."*

*Moreover, the guidelines state, "Where significant controversy exists on a subject, for example, for many topics that are studied within the foundations of physics communities, AJP is usually not an appropriate venue. Disagreements among experts should be settled within the research literature."*

*Unfortunately, AJP can only publish about 15% of the manuscripts it receives, so we have to be very selective in choosing the articles that will be of the most interest to our readers.*

*Therefore, I regret to inform you that we will not pursue the publication of your manuscript.*

**Foundations of Physics:** be4c910b-35a2-4c12-ad11-a3240789802d

**Editor:** *Your manuscript entitled "Positivism in Newtonian Mechanics: The Ousia or a Historical Liability?" has now been assessed. If there are any reviewer comments on your manuscript, please find them below.*

*Regrettably, the above submission has been rejected for publication in Foundations of Physics.*

**The British Journal for the Philosophy of Science:** BJPS-2023-238

Submission date: 2023-07-13

**Editor:** *We regret to inform you that we have decided not to accept your paper for publication. As you will appreciate, the Journal receives a large number of high quality submissions every year, and consequently competition for space in the journal is intense. We are sorry not to be able to provide more detailed feedback, but this enables us to give you a speedier response.*

**Studies in History and Philosophy of Science:** SHPS-D-23-00280

Submission date: 2023-07-28

**Editor:** *Based on our associate editor's reading of your manuscript, we have decided to reject it in its current form as it does not meet the required quality standards for scholarship in this journal as it stands. However, we see promise in aspects of the article and would like to invite you to resubmit a reworked article focusing on the questions of determinism/indeterminism in classical physics.*

*The term "positivism" has been used in many ways, but the close ties to logical positivism in the philosophy of science risks obscuring the independent points of your manuscript and makes it difficult to secure the right referees for your manuscript. Should you decide to resubmit, please pay particular attention to rewriting the abstract in order to focus more clearly on what is now the middle paragraph "Four scientific topics, showing two different types of indeterminacy, are introduced and briefly reviewed: chaotic systems, turbulence, fluid transport in porous media, and hydromechanics of fractures. It is proposed that determinism is only meaningful in epistemic level, the first two examples are ontically indeterministic, the last two examples are epistemically indeterministic".*

**HOPOS: The Journal of the International Society for the History of Philosophy of Science:** 2143

Submission date: 2023-08-20

**Editor:** *Thank you for submitting your paper, "Positivism in Newtonian Mechanics: The Ousia or a Historical Liability?," for consideration for publication. This paper is making a systematic argument in the philosophy of science, and its main aim is not a historical or social analysis of the philosophy of science. As such, the paper is outside the scope of our journal. We recommend that you seek another publication venue.*

**Continental Philosophy Review:** MAWO-D-23-00161

Submission date: 2023-08-31

**Editor:** *Thank you for submitting your manuscript "Positivism in Newtonian Mechanics: The Ousia or a Historical Liability?" to Continental Philosophy Review. We regret to inform you that we are not in a position to pursue the publication of this submission. Unfortunately we are not able to provide reviewer comments in this instance.*

**Journal of Applied Philosophy:** JAPP-2023-243-OA

Submission date: 2023-09-21

**Editor:** *Thank you for submitting your paper to the Journal of Applied Philosophy. I'm sorry to tell you that the decision is against accepting your paper. We receive a large number of high quality submissions every year, thus competition for space is intense. After careful consideration, we have decided on editorial grounds not to send this paper for external review.*

*Unfortunately, due to the volume of submissions we receive, we are unable to provide a detailed explanation on why your paper has not been taken forward into our review process. We hope that you find a journal which is a better fit.*

**Tradition and Discovery: The Polanyi Society Journal**

Submission date: 2023-11-17

**Editor's Comment**
*The reviewer agrees that is a worthwhile task. At the same time, the reviewer does not find the argument persuasive and identifies four weak areas.*
*I have also read through the submission again and see your main line of argument as something like this: Newtonian mechanics is positivistic at its root and that since positivism is a discredited philosophical orientation, then we may need to rethink some assumptions behind Newtonian mechanics, or at least the idea that what we can know by means of Newtonian mechanics (epistemological issues) exactly corresponds to the nature of reality (ontic issues). I think that is a promising line of argument, but since physics and engineering is not my field, I am not competent to assess the technical details you include here.*

**Authors' Response**
I appreciate the reviewer's time and effort on this manuscript. I have incorporated additional discussions to the manuscript to clarify the points raised by the reviewer. I have also discussed these points in detail in the following. Your insightful reflection on the manuscript captures its essence well, and I am delighted that you have found it a promising line of argument.

**Reviewer's Comment #1**
*(1) The author notes that chaotic systems are deterministic but unpredictable, but seems to reject that by citing Bishop on p. 11.*

**Authors' Response #1**
While I have presented the notion of "deterministic, but unpredictable", commonly employed for chaotic systems by philosophers of science including Bishop (2017), I have <u>not</u> embraced this apologia in this manuscript, due to the ontic indeterminacy of chaotic systems, discussed in section 4.2, and also its philosophical inconsistency, discussed in section 6. In short, I do not consider the notion of "deterministic, but unpredictable" a valid interpretation for chaotic systems, since even obtaining the analytical solution for a chaotic equation, such as the one presented in Equation (2), does not result in a deterministic answer for that equation (one output for one input). Regarding this point, I have included further discussions on Navier-Stokes equations for turbulence modeling in the endnote 35.

- <u>Endnote 35, Line 3</u>: Let's suppose even an analytical solution for Navier-Stokes equations is proposed. The turbulence still remains indeterministic, much like the analytical solution of chaotic Equation (1), presented in Equation (2), which cannot result in a deterministic answer (one output for one input).

**Reviewer's Comment #2**
*(2) The author needs to recognize that significant digits (p. 13) are meaningful only when trying to quantify physical phenomena.*

**Authors' Response #2**
The reviewer's attention to this point is appreciated. In this context, in fact, another interpretation of determinism has been indirectly questioned. According to the deterministic-positivistic assumptions inherent in the natural scientific worldview, the "laws" of nature, such as Newton's laws of motion, are "unlimitedly" accurate. This is, however, only an assumption, which could be questionable for a natural reality. For example, Newton's laws of motion are not necessarily accurate interpretations of a quantum-level or a relativity-physics phenomenon. A new endnote is added to the text to further clarify this point.

- Endnote 32: According to our deterministic presumptions, an "accurate" equation, e.g., Newton's laws of motion, is "unlimitedly" accurate. This is, of course, merely an ontological assumption within a positivistic worldview/paradigm, which could be questioned for a natural reality. It is, at least, known that in the quantum level or in the realm of relativity physics, defending such accuracies becomes a challenging task.

### Reviewer's Comment #3
*(3) The example of coin-tossing seems to equate unpredictability with indeterminism and that is not really the case.*

### Authors' Response #3
Coin tossing is an interesting example in determinism literature. It is a traditional example on randomness, but it has a detailed mathematical literature discussing its determinability (see e.g., Physics Reports 469 (2008) 59–92). It, however, aligns well with what I have termed "ontic indeterminacy" in this manuscript.

The abstract of the cited reference starts with "[t]he dynamics of the tossed coin can be described by deterministic equations of motion…", but it notably ends with "…practically any uncertainty in initial conditions can lead to the uncertainty of the results of tossing". In other words, similar to a chaotic system, it is assumed to be "deterministic, but unpredictable", specifically, as the authors of this reference describe in the Conclusions section that "coin bouncing on the floor" leads to "transient chaotic behaviour"! When the inherent uncertainties associated with a human subject tossing a coin are also considered, it becomes evident that the phenomenon is inherently indeterministic. To avoid any potential misunderstanding for the readership, a new endnote is added to the text to clarify that coin tossing is proposed as an example for an indeterministic phenomenon.

- Endnote 27: Coin tossing is proposed as an example for a stochastic phenomenon in this text.

### Reviewer's Comment #4
*(4) The author excludes thermodynamics early in the paper, but doing so excludes any statistical argument for indeterminism.*

### Authors' Response #4
I have excluded thermodynamics and entropy-related phenomena since, as Polanyi (PK, 412-415) has beautifully demonstrated, they bring other types of obstacles in front of determinism and positivism, which is beyond the scope of this manuscript. Statistical analysis serves as a powerful tool in many scientific and engineering fields, which are, in essence, unrelated to thermodynamics. The reference books on statistical methods and analyses (e.g., D.C. Montgomery, *Design and analysis of experiments*, John Wiley & Sons, 2009) typically do not even mention thermodynamics.
Moreover, while statistical thermodynamics is a well-established branch of thermodynamics, the science of thermodynamics is not, by any means, limited to statistical thermodynamics. Many reference books on thermodynamics, particularly those designed for undergraduate courses, either briefly discuss statistical thermodynamics, or exclude it entirely. In short, thermodynamics and statistical analyses are two independent fields of studies. At any rate, I have added a new endnote to the text to clarify the rationale behind excluding thermodynamics in the manuscript.

- <u>Endnote 4</u>: Such phenomena could bring other types of obstacles in front of determinism and positivism, as discussed by Polanyi (1962, 412-415).

## Editor's Final Comment

Response date: 2024-05-22

*I regret to inform you that we will not be able to publish your paper in TAD. We have not gotten any response from a second reviewer and cannot find a replacement. Furthermore, while the general content seems interesting, the connections with Polanyi's work are rather tenuous.*